\documentclass[a4paper,twocolumn,accepted=2021-06-15]{quantumarticle}
\pdfoutput=1
\usepackage[english]{babel} 
\usepackage[T1]{fontenc}
\usepackage[latin9]{inputenc}
\usepackage{color}
\usepackage{xcolor}
\usepackage{placeins}
\usepackage{amsmath,amsfonts,amsthm}
\newtheorem{proposition}{Proposition}
\newtheorem{conjecture}[proposition]{Conjecture}

\usepackage{url}

\usepackage{breakurl}

\usepackage{graphicx}
\usepackage{tabularx}
\usepackage[numbers,sort&compress]{natbib}
\usepackage{braket}
\usepackage{amssymb}
\newcommand{\R}{\mathbb R}
\DeclareMathOperator{\tr}{tr}
\DeclareMathOperator{\Prob}{Prob}
\DeclareMathOperator{\sgn}{sign}
\DeclareMathOperator{\diag}{diag}
\DeclareMathOperator{\opt}{opt}
\usepackage[unicode, breaklinks]{hyperref}
\usepackage[ruled,vlined]{algorithm2e}
\usepackage{setspace}

\usepackage{tikz}
\def\checkmark{\tikz\fill[scale=0.4](0,.35) -- (.25,0) -- (1,.7) -- (.25,.15) -- cycle;} 

\makeatother

\usepackage{babel}

\begin{document}

\title{Warm-starting quantum optimization}

\author{Daniel J. Egger}
\orcid{0000-0002-5523-9807}
\email{deg@zurich.ibm.com}
\affiliation{IBM Quantum, IBM Research -- Zurich, S\"aumerstrasse 4, 8803 R\"uschlikon, Switzerland}
\author{Jakub Mare\v{c}ek}
\orcid{0000-0003-0839-0691}
\email{jakub.marecek@fel.cvut.cz}
\affiliation{Czech Technical University, Karlovo nam.~13, Prague 2, the Czech Republic}
\author{Stefan Woerner}
\orcid{0000-0002-5945-4707}
\email{wor@zurich.ibm.com}
\affiliation{IBM Quantum, IBM Research -- Zurich, S\"aumerstrasse 4, 8803 R\"uschlikon, Switzerland}

\begin{abstract}
There is an increasing interest in quantum algorithms for problems of integer programming and combinatorial optimization. 
Classical solvers for such problems employ relaxations, which replace binary variables with continuous ones, for instance in the form of  higher-dimensional matrix-valued problems (semidefinite programming).
Under the Unique Games Conjecture, these relaxations often provide the best performance ratios available classically in polynomial time.
Here, we discuss how to warm-start quantum optimization with an initial state corresponding to the solution of a relaxation of a combinatorial optimization problem and how to analyze properties of the associated quantum algorithms.
In particular, this allows the quantum algorithm to inherit the performance guarantees of the classical algorithm.
We illustrate this in the context of portfolio optimization, where our results indicate that warm-starting the Quantum Approximate Optimization Algorithm (QAOA) is particularly beneficial at low depth.
Likewise, Recursive QAOA for MAXCUT problems shows a systematic increase in the size of the obtained cut for fully connected graphs with random weights, when Goemans-Williamson randomized rounding is utilized in a warm start. 
It is straightforward to apply the same ideas to other randomized-rounding schemes and optimization problems.
\end{abstract}

\maketitle

\section{Introduction}

Gate-based quantum computers are expected to help solve problems in quantum chemistry \cite{Moll2018, Kandala2018, Ganzhorn2019}, machine learning \cite{Biamonte2017, Havlicek2019}, financial simulation \cite{Egger2020, Woerner2019, Rebentrost2018, Stamatopoulos2019, Martin2019, Orus2019, Egger2019, Vazquez2020} and combinatorial optimization \cite{Braine2019, Barkoutsos2020}.
The quantum approximate optimization algorithm (QAOA) \cite{Farhi2014, Farhi2014b, Yang2017}, inspired by a Trotterization of adiabatic quantum computing \cite{Johnson2011, Mbeng2019, Juenger2019}, runs on gate-based quantum computers \cite{Barends2016, Willsch2020}.
This algorithm encodes a combinatorial optimization problem in a Hamiltonian $\hat H_C$ whose ground state is the optimum solution.
The QAOA first creates an initial state which is the ground state of a mixer Hamiltonian $\hat H_M$.
A common choice of $\hat H_M$ and initial state is $-\sum_{i=0}^{n-1}\hat X_i$ and $\ket{+}^{\otimes n}$, respectively.
Next, in a QAOA with depth $p$, a quantum circuit applies $\exp{(-i\beta_k\hat H_M)}\exp{(-i\gamma_k\hat H_C)}$ at each layer $k=1,...,p$ to create a trial state $\ket{\psi(\boldsymbol{\beta}, \boldsymbol{\gamma})}$.
A classical optimizer seeks the optimal values of $\boldsymbol{\beta}$ and $\boldsymbol{\gamma}$ to create a trial state which minimizes the energy of the Hamiltonian $\hat H_C$.
This algorithm has lacked theoretical guarantees on its performance ratio and for certain problem instances of MAXCUT it cannot, with constant depth, outperform the classical Goemans-Williamson randomized rounding approximation \cite{Bravyi2019, Crooks2018}.

Recent work has improved the original QAOA, for instance, by aggregating only the best sampled candidate solutions \cite{Barkoutsos2020} and carefully choosing the mixer operator to improve convergence \cite{Farhi2017,Hadfield2019,zhu2020,Wang2020}, empirically.
Reinforcement learning \cite{Khairy2020, Wauters2020}, multi-start methods \cite{Shaydulin2019}, and local optimization \cite{Shaydulin2019b} help navigate the QAOA optimization landscape.
Algorithms such as the Hamiltonian Variational Ansatz produce optimization landscapes that are easier to navigate \cite{wiersema2020}.
Furthermore, optimal $\boldsymbol{\beta}$ and $\boldsymbol{\gamma}$ values concentrate on all typical instances generated by some reasonable distributions which may allow optimization strategies with fewer calls to the quantum computer \cite{Brandao2018}.
Certain local classical algorithms match the performance of QAOA for Ising-like cost functions with multi-spin interactions \cite{Hastings2019} which has motivated the development of Recursive-QAOA (RQAOA) \cite{Bravyi2019, Bravyi2020b}.
RQAOA iteratively reduces the problem size and outperforms QAOA on certain forms of Ising Hamiltonians \cite{Bravyi2019}.
Implementing QAOA on noisy quantum hardware is challenging as the number of gates can be high for current gate fidelities \cite{Alam2019,Akshay2020}.
The circuits become especially deep when large $p$ is required or when the native hardware connectivity does not match the problem structure, thence requiring SWAP gates \cite{Harrigan2021}.
Therefore, in the near term, quantum computers will most likely only run low-depth QAOA.
Low-depth QAOA results are improved by robust control \cite{Dong2019} and by mapping $\boldsymbol{\beta}$ and $\boldsymbol{\gamma}$ to parameters of the control pulses \cite{Lacroix2020, Earnest2021}, a method available to cloud-based quantum computers \cite{Gokhale2020} with pulse-level control \cite{Mckay2018,Alexander2020}.

Recently, there has been substantial progress \cite{majumdar2020recent} in the study of continuous relaxations of NP-Hard combinatorial  optimization problems. 
The best-known continuous relaxations of MAXCUT and many other problems take the form of semidefinite programs \cite{anjos2011handbook}.
These can be solved efficiently both in theoretical models of computation \cite{blum2012complexity}, where a real-number arithmetic operation can be performed in unit time, and in practice\footnote{Notice that the analysis of  \cite{porkolab1997complexity} shows the situation is less trivial in the Turing machine and one may need to consider the dimension or the number of constraints a constant, as it often is for a particular relaxation.} \cite{yurtsever2019scalable}.
Subsequently, the solution of a continuous relaxation of a combinatorial optimization problem is transformed into a good solution of the discrete-valued  problem by randomized rounding \cite{Raghavan1987}.
For instance, the celebrated Goemans-Williamson (GW) random hyperplane rounding \cite{Goemans1995, Karloff1999} for MAXCUT finds cuts whose expected value is an $\alpha$ fraction of the global optimum, for $0.87856<\alpha<0.87857$, with the expectation over the randomization in the rounding procedure.
The Unique Games Conjecture \cite{khot2007optimal,KhotSurvey,khot2015unique}, introduced in Appendix~\ref{sec:ucg}, suggests that GW randomized rounding has the best possible
polynomial-time performance on MAXCUT.

Our work is motivated by the desire to provide at least as good guarantees for QAOA as there are for classical approximations.
For example, we show that for MAXCUT a warm-start can preserve the GW approximation ratio at any depth $p$~\cite{Marwaha2021}. Further, if the Unique Games Conjecture were to be false even stronger guarantees may be available, improving upon those for randomized rounding.
In simulations, our variant of QAOA consistently performs as well as the GW algorithm or better.

We discuss how to warm-start quantum optimization in Sec.~\ref{sec:warm_start}.
We explore warm-starting QAOA (WS-QAOA) numerically in Sec.~\ref{sec:simulations} by relaxing Quadratic Unconstrained Binary Optimization problems to continuous ones which provide QAOA with a good initial solution.
In Sec.~\ref{sec:max_cut} we use the GW algorithm \cite{Goemans1995} to warm-start RQAOA.
We discuss our results and conclude in Sec.~\ref{sec:conclusion}.

\section{Warm-start Quantum Opti\-mi\-za\-tion}

\subsection{Preliminaries}

Quadratic Unconstrained Binary Optimization (QUBO) has been studied in Combinatorial Optimization since the 1960s \cite{hammer1968}. A common formulation is
\begin{align}\label{eqn:qubo}
    \min_{x\in\{0,1\}^n} x^T\Sigma x+\mu^Tx, \tag{QUBO}
\end{align}
where $x$ is a vector of $n$ binary decision variables, $\Sigma\in\mathbb{R}^{n\times n}$ a symmetric matrix, and $\mu\in\mathbb{R}^n$ a vector.
Since for binary variables $x_i^2=x_i$, $\mu$ can be added to the diagonal of $\Sigma$, and in the following, we only add $\mu$ when it simplifies the notation in the given context.
Considering that any mixed-integer linear program can be encoded in a QUBO \cite{LASSERRE2016}, QUBO is NP-Hard. Indeed, even checking local optimality is NP-Hard \cite{pardalos1988checking}, and hence only very special cases
\cite{allemand2001polynomial,hladik2019new}
can be solved in polynomial time. 

If $\Sigma$ is positive semidefinite, the trivial continuous relaxation of QUBO
\begin{align}\label{eqn:relaxed}
    \min_{x\in[0, 1]^n} x^T\Sigma x,\tag{QP}
\end{align}
is a convex quadratic program and the optimal solution $c^*$ of the continuous relaxation is easily obtainable with classical optimizers \cite{gondzio2006solving}. 

If $\Sigma$ is not positive semidefinite, one can apply the well-known recipe \cite{poljak1995recipe} to obtain another continuous-valued relaxation, known as semidefinite programming (SDP):
\begin{align}
    \max_{Y \in \mathbb{S}^{n}} \tr(\Sigma Y) \tag{SDP} \label{eq:SDP} \\
    \diag(Y) = e \notag \\
    Y \succeq 0, \notag 
\end{align}
where $\mathbb{S}^{n \times n}$ denotes the set of $n \times n$ symmetric matrices, $e$ is an $n$-vector of ones, and $Y \succeq 0$ denotes that $Y$ must be positive semidefinite.
Given the optimal solution $Y^*$ to \eqref{eq:SDP}, there exist several approaches to generating solutions of the corresponding \eqref{eqn:qubo}, often with approximation guarantees, as discussed later in this section and Appendix \ref{sec:gw}.
A classical laptop can solve instances of \eqref{eq:SDP} relaxations of QUBO, where $\Sigma$ has $10^{13}$ entries \cite{yurtsever2019scalable}.
Furthermore, quantum computers offer the prospect of  some speed-ups in solving SDPs \cite{van2020quantum,brandao2019quantum}, although recent quantum-inspired algorithms for SDPs may reduce the potential speedup \cite{chia2020quantum}.

\subsection{Continuous warm-start QAOA \label{sec:warm_start}}

\begin{figure}
    \centering
    \includegraphics[width=0.95\columnwidth]{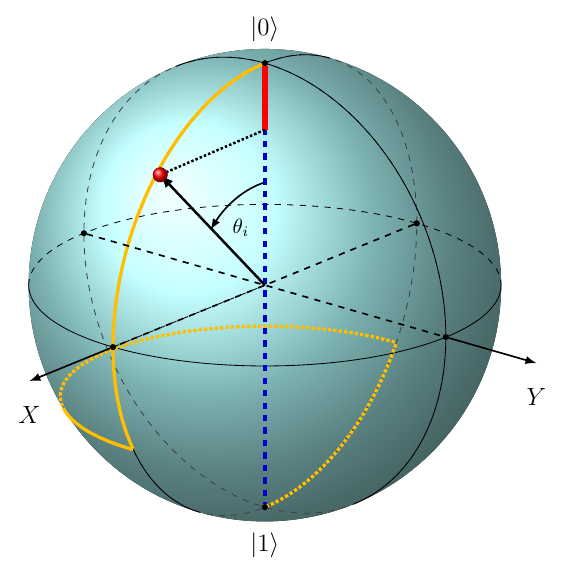}
    \caption{Intial state $\hat R_Y(\theta_i)\ket{0}$ (red point) on the Bloch sphere.
    The dashed blue line from $\ket{0}$ to $\ket{1}$ is the interval $[0,1]$. 
    The red line is the solution $c_i^*$ to the relaxed problem.
    The regularization parameter $\varepsilon$ restricts the range of allowed rotation angles.
    The orange path shows the evolution of the quantum state starting at $\ket{0}$ for $c_i^*=0$, $\varepsilon=0.25$ and $\beta=\pi/2$ with the rotations $\hat R_Y(-\theta_i)\hat R_Z(-2\beta)\hat R_Y(2\theta_i)$.}
    \label{fig:bloch}
\end{figure}

The solutions of either continuous-valued relaxation (\ref{eqn:relaxed} or \ref{eq:SDP}) can be used to initialize  quantum-classical hybrid algorithms, which is known as  warm-starting them~\cite{gondzio1998warm}.
In particular, we focus on warm-starting QAOA.

In QAOA, each decision variable $x_i$ of the discrete optimization problem corresponds to a qubit by the relation $x_i=(1-z_i)/2$. 
Each $z_i$ is replaced by a spin operator $\hat Z_i$ to transform the cost function to a cost Hamiltonian $\hat H_C$ \cite{Lucas2014,lodewijks2019mapping}.
Note that the final measurement in QAOA can be considered as a randomized rounding.
In the simplest variant of WS-QAOA, we replace the initial equal superposition state $\ket{+}^{\otimes n}$ with a state
\begin{align}
    \ket{\phi^*}=\bigotimes_{i=0}^{n-1} \hat R_Y(\theta_i)\ket{0}_n,
    \label{eq:relaxedstate}
\end{align}
which corresponds to the solution $c^*$ of the relaxed Problem~\eqref{eqn:relaxed}.
Here, $\hat R_Y(\theta_i)$ is a rotation around the Y-axis of qubit $i$ with angle $\theta_i=2\arcsin\left(\sqrt{c_i^*}\right)$ and $c_i^*\in[0,1]$ is the $i$-th coordinate of the optimum of the continuous-valued relaxation \eqref{eqn:relaxed}.
The probability to measure $\ket{1}$ in qubit $i$ is thus $c^*_i$, see the geometric representation in Fig.~\ref{fig:bloch}.

\begin{figure}
    \centering
    \includegraphics[width=\columnwidth]{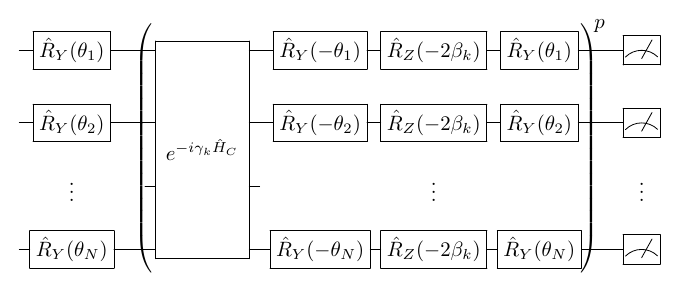}
    \caption{
    Quantum circuit for WS-QAOA. 
    The first $\hat R_Y$ rotations prepare the initial state $\ket{\phi^*}$.
    The mixer operator, i.e. $\hat R_Y(\theta_i)\hat R_Z(-2\beta_k)\hat R_Y(-\theta_i)$, is applied after the time-evolved problem Hamiltonian $\hat H_C$.
    }
    \label{fig:warm_circuit}
\end{figure}

We also replace the mixer Hamiltonian $\hat H_M=-\sum_{i=0}^{n-1}\hat X_i$ with $\hat H_M^{(ws)}=\sum_{i=0}^{n-1}\hat H_{M, i}^{(ws)}$ where
\begin{align}
    \hat H_{M, i}^{(ws)}=
    \begin{pmatrix}
        2c_i^*-1 & -2\sqrt{c_i^*(1-c_i^*)} \\
        -2\sqrt{c_i^*(1-c_i^*)} & 1-2c_i^*
    \end{pmatrix}
    \label{eq:new_mixer}
\end{align}
and has $\hat R_Y(\theta_i)\ket{0}$ as ground state with eigenvalue of $-1$.
The ground state of $\hat H^{(ws)}_M$ is thus $\ket{\phi^*}$ with eigenvalue $-n$~\cite{Farhi2014}.
Therefore, WS-QAOA applies at layer $k$ a mixing gate which is given by the time-evolved mixing Hamiltonian $\exp(-i\beta_k\hat H_M^{(ws)})$, see Fig.~\ref{fig:warm_circuit}.
Since $\hat H_{M, i}^{(ws)}=-\sin(\theta_i)\hat X-\cos(\theta_i)\hat Z$ the time-evolved mixing Hamiltonian is a rotation around the axis $\vec{n}=[-\sin(\theta_i), 0, -\cos(\theta_i)]$ on the Bloch-sphere of qubit $i$ and can be implemented using the single-qubit rotations $\hat R_Y(\theta_i)\hat R_Z(-2\beta)\hat R_Y(-\theta_i)$.

If a coordinate in the optimal solution of a continuous relaxation is $c_i^*=0$ or $c_i^*=1$, qubit $i$ would be initialized in state $\ket{0}$ or $\ket{1}$, respectively.
In such cases, the qubit will remain in its initial state throughout the QAOA optimization when $\hat H_C$ contains only $\hat Z_i\hat Z_j$ and identity spin-operators.
This creates a reachability issue when the optimal continuous and discrete solutions do not overlap, i.e., $d_i^*=1$ and $c_i^*=0$ or $d_i^*=0$ and $c_i^*=1$, where $d^*$ is the solution to the \eqref{eqn:qubo}.

To mitigate this effect, we introduce a variant of WS-QAOA that utilizes a regularization parameter $\varepsilon\in[0,0.5]$ and changes the rotation angle creating the initial state according to
\begin{alignat*}{2}
\label{eq:usevarepsilon}
    \theta_i=&\,2\arcsin\left(\sqrt{c_i^*}\right) &&\quad\text{if}\quad c_i^*\in[\varepsilon, 1-\varepsilon], \\
    \theta_i=&\,2\arcsin\left(\sqrt{\varepsilon}\right) &&\quad\text{if}\quad c_i^*\leq \varepsilon, \\
    \theta_i=&\,2\arcsin\left(\sqrt{1-\varepsilon}\right) && \quad\text{if}\quad c_i^*\geq 1-\varepsilon.
\end{alignat*}
The mixer Hamiltonian is adjusted accordingly.
The parameter $\varepsilon$ provides a continuous mapping between WS-QAOA and standard QAOA since at $\varepsilon=0.5$ the initial state is the equal superposition state and the mixer Hamiltonian is the $X$ operator.
If all $c_i^* \in (0, 1)$ or $\varepsilon > 0$, WS-QAOA converges to the optimal solution of \eqref{eqn:qubo} as the depth $p$ approaches infinity as does standard QAOA~\cite{Farhi2014}.
This directly follows from the adiabatic theorem and the fact that we start in the ground state of the mixer which overlaps with all computational basis states including the optimal solution.
For large enough $p$, (WS-)QAOA therefore reproduces the adiabatic evolution transforming the ground state of the mixer into the ground state of $\hat H_C$.
The speed of the adiabatic evolution is limited by the spectral gap of the intermediate Hamiltonians.
If the evolution is too fast transitions to excited states occur which may not result in the optimal solution. 
The speed of the evolution can be related to $p$, where a slow evolution, i.e., longer total evolution time, implies a larger $p$.
The idea of (WS-)QAOA is to speed-up this evolution by optimizing the parameters instead of following a fixed annealing schedule.

\begin{table*}[tb!]
    \centering
        \caption{An overview of design choices in warm-starting a quantum optimization algorithm for \eqref{eqn:qubo}.
        Under ``What to round?'', columns are ordered left to right to suggest the increasing strength of the relaxations, although this is necessarily fraught in the case of hierarchies of relaxations \cite{lasserre2001global,lasserre2006convergent,ghaddar2011second}, where one column represents a potentially infinite number of relaxations.
        Similarly, under ``How to round?'', we order the options approximately by their performance.
}
    \label{tab:designchoices}
    \begin{tabular}{l|ccccc|cc|ccccc} \hline\hline
Variant & \multicolumn{5}{l|}{What to round?} & \multicolumn{2}{l|}{When to round?} & \multicolumn{5}{l}{How to round?}\\ \hline
         & \rotatebox[origin=c]{90}{ \eqref{eqn:relaxed} Relaxation } & \rotatebox[origin=c]{90}{SOCP Relaxations \cite{ghaddar2011second}} & \rotatebox[origin=c]{90}{\eqref{eq:SDP} Relaxation \cite{poljak1995recipe,Goemans1995,charikar2004maximizing}} \; & 
         \rotatebox[origin=c]{90}{Entropy-penalized SDP \cite{krechetov2019entropy}}
         \rotatebox[origin=c]{90}{Sparse Moment SDP \cite{lasserre2006convergent}} & \rotatebox[origin=c]{90}{Moment SDP \cite{lasserre2001global}} & \rotatebox[origin=c]{90}{Classical pre-processing} & \rotatebox[origin=c]{90}{Quantum circuit (QC)} &
         \rotatebox[origin=c]{90}{0.5-Approximation \cite{sahni1976p,mitzenmacher2017probability}} & 
         \rotatebox[origin=c]{90}{Random hyperplane \cite{Goemans1995,charikar2004maximizing}} & \rotatebox[origin=c]{90}{Iterative \cite{abbasi2020sticky,eldan2019krivine}} & \rotatebox[origin=c]{90}{Iterative \cite{morgenstern2019fair}} & 
         \rotatebox[origin=c]{90}{Measurement in a QC} \; \\ \hline
WS-QAOA & \checkmark & & & & & & \checkmark & & & & & \checkmark \\
WS-RQAOA & & & \checkmark & & & \checkmark &  & &  \checkmark & & & \\ \hline\hline
    \end{tabular}
\end{table*}

\subsection{Rounded warm-start QAOA}
\label{sec:rounded}

Further variants of WS-QAOA randomly round the optimum of the continuous-valued relaxation before using it as the initial state.
This is appealing to quantum hardware with limited qubit numbers as even for convex relaxations in dimensions that scale super-linearly with the number $n$ of binary variables in \eqref{eqn:qubo}, such as \eqref{eq:SDP} with dimension $n(n+1)/2$, the representation of the rounded solution to \eqref{eqn:qubo} requires only $O(n)$ qubits.
Two notable examples are the random-hyperplane rounding of SDP relaxations for MAXCUT \cite{Goemans1995}, see Appendix \ref{sec:gw},  and iterative rounding of SDP relaxations for a wider variety of problems, see Appendix \ref{sec:an}.
Both of these examples provide initial states that already have the best approximation guarantee available classically in polynomial time.

We now elaborate on the example of Goemans-Williamson random-hyperplane rounding of \eqref{eq:SDP}.
For a given GW cut we generate an initial state using $Y$-rotations with a $\varepsilon\in(0, 0.5)$, as discussed in Sec.~\ref{sec:warm_start}.
To warm-start QAOA such that we can 
retain the GW bound on MAXCUT, we wish create a quantum circuit that can both represent solutions of the random-hyperplane rounding as well as deviate from them.
We therefore modify the mixer such that its time-evolution is $\hat R_Y(-\theta_i) \hat R_Z(-2\beta) \hat R_Y(\theta_i)$ instead of $\hat R_Y(\theta_i) \hat R_Z(-2\beta) \hat R_Y(-\theta_i)$, i.e., we multiply the off-diagonal elements in \eqref{eq:new_mixer} by $-1$.
With this modification, the value of the regularization parameter $\varepsilon$ can be set to $0.25$ to generate states that differ from the GW rounding as well as retain it by choosing $\beta_1=\pi/2$ and $\gamma_1=0$.
At these values the depth-one variational form reduces to
\begin{eqnarray}
\hat R_Y(-\theta_i) \hat R_Z(-\pi) \hat R_Y(2\theta_i) \ket{0},
\end{eqnarray}
for each qubit, and creates the states $-i \ket{1}$ and $-i \ket{0}$ when $c_i^*=\varepsilon$ and $1-\varepsilon$, respectively.
Thus, the variational form can recover the solution given by the GW rounding, considering that $z$ and $1 - z$ represent the same cut, see the orange path in Fig.~\ref{fig:bloch} as example.
Therefore, WS-QAOA is at least as good as GW rounding.
This adjustment also comes with a drawback. Since the prepared initial state is no longer an eigenstate of the mixer (otherwise we would not be able to deviate from it) we cannot use the same arguments as in \cite{Farhi2014} to derive the convergence of the algorithm to the global optimum with increasing depth $p$.
We will analyze this numerically in Sec.~\ref{sec:max_cut}.

Notice that measuring an initial state provided by a randomized rounding of the semidefinite programming relaxation \eqref{eq:SDP} yields the best approximation guarantees available classically in polynomial time under the Unique Games Conjecture \cite{khot2007optimal,KhotSurvey,khot2015unique}.
Therefore, any quantum circuit that preserves or improves the performance ratio would preserve or improve the overall performance guarantees.

\begin{figure}
    \centering
    \includegraphics[width=\columnwidth]{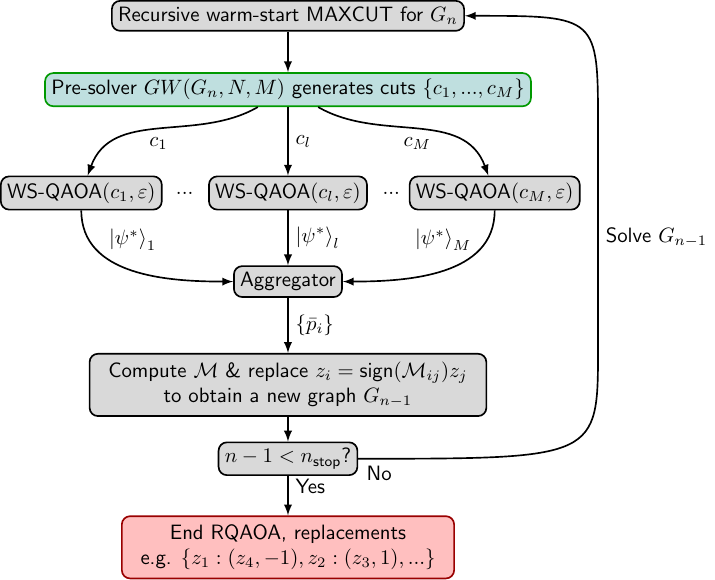}
    \caption{WS-RQAOA for MAXCUT.
    At each iteration we run several WS-QAOAs that are initialized with different solutions from the GW randomized rounding.
    The resulting samples are aggregated to compute the combined correlation matrix $\mathcal{M}$ needed by RQAOA to eliminate a decision variable.}
    \label{fig:max_cut_algo}
\end{figure}

Rounding in the classical pre-processing readily leads to the warm-started recursive QAOA (WS-RQAOA), illustrated in Fig.~\ref{fig:max_cut_algo} and demonstrated in Sec.~\ref{sec:max_cut}.
For MAXCUT of a graph $G_n$, we leverage a GW pre-solver $GW(G_n,N,M)$ to generate $N$ good cuts of which we retain the $M<N$ best unique cuts.
These $M$ cuts therefore initialize $M$ WS-QAOA optimizers with $\varepsilon\in\,(0,0.5)$.
Each QAOA solver produces an optimized variational state $\ket{\psi^*}_l=\sum_{i=0}^{2^n-1}\alpha_{il}\ket{i}$ for $l=1,...,M$.
We then aggregate these $M$ variational states by averaging the probability of sampling each bit-string $\ket{i}$, i.e. $\bar p_i=M^{-1}\sum_{l=1}^M|\alpha_{il}|^2$, and use these average probabilities to create the correlation matrix $\mathcal{M}$ needed by RQAOA \cite{Bravyi2019, Bravyi2020b}, see Appendix~\ref{sec:rqaoa} and \ref{sec:efficient_rqaoa}.
At each iteration, RQAOA removes one decision variable $z_i$ from the problem by replacing it with $\text{sign}(\mathcal{M}_{ij})z_j$, where $(i,j) = \arg\max_{(i,j)} |\mathcal{M}_{ij}|$.
This generates a new MAXCUT problem with a new graph $G_{n-1}$, see Appendix \ref{sec:max_cut_reduction}, for which we repeat this procedure, illustrated in Fig.~\ref{fig:max_cut_algo}, until the reduced graph reaches a certain size $n_\text{stop}$.
The graph $G_{n_\text{stop}}$ is solved by diagonalizing the Hamiltonian $\hat H_C$ or by applying classical optimizers.

\subsection{Further variants of warm-starting quantum optimization}

In Sec.~\ref{sec:warm_start} and \ref{sec:rounded}, we gave first examples of how to warm-start QAOA using a continuous relaxation and a randomized rounding. 
The key algorithm-design questions in warm-starting quantum optimization are: what to round, when to round it, and how to round it.
For each of these questions, there are multiple options available, as suggested in the previous discussion and summarized in Tab.~\ref{tab:designchoices}.

First, there are many options for what to round, outside of the \eqref{eqn:relaxed} relaxation and the \eqref{eq:SDP} relaxation.
For example, the \eqref{eqn:relaxed} relaxation can be seen as a second-order cone programming (SOCP) relaxation, and could be strengthened iteratively \cite{ghaddar2011second}, until its objective-function value coincides with the objective-function value of the non-convex \eqref{eqn:qubo}, albeit at the cost of an exponential growth of the relaxation.
Similarly, one could strengthen the \eqref{eq:SDP} relaxation either by using an entropy-penalizing term    \cite{krechetov2019entropy} or by using the Moment/SOS hierarchy \cite{lasserre2001global} and its sparse variant \cite{lasserre2006convergent}.

Second, there are two options for when to round: 
either in the classical pre-processing --- within the initial state preparation which leads to the WS-RQAOA discussed in Sec.~\ref{sec:rounded} on the example of the \eqref{eq:SDP} relaxation --- or within the quantum circuit.
In its simplest form, the latter can be a quantum measurement, as discussed in Sec.~\ref{sec:warm_start} on the example of the \eqref{eqn:relaxed} relaxation.

Third, there are several options for the rounding procedure. Even the simplest rounding mechanisms often perform well: on MAXCUT, for example, disregarding the relaxation and coordinate-wise assigning a value uniformly at random achieves a 0.5 approximation ratio \cite{sahni1976p} and can be derandomized \cite[Chapter 6]{mitzenmacher2017probability}.
The random-hyperplane rounding of GW \cite{Goemans1995}, as explained in Appendix \ref{sec:gw}, 
improves the performance ratio on MAXCUT to $\alpha ={\frac {2}{\pi }}\min _{0\leq \theta \leq \pi }{\frac {\theta }{1-\cos \theta }}\approx 0.878$.
The same ratio can also be  obtained with an iterative procedure that rounds coordinates that are close to being integral to integers \cite{abbasi2020sticky,eldan2019krivine} and removes them from further processing\footnote{Notice that when the WS-QAOA leverages both a classical optimization routine with a classical conditional statement and a noisy quantum computer, such a rounding  procedure can also be considered.}, as explained in Appendix \ref{sec:an}.
Plausibly, the same ratio could also be achieved with a number of novel and very different iterative procedures, such as  \cite{morgenstern2019fair}.

\subsection{Discussion of warm-starting quantum optimization}

On noisy quantum hardware it seems appealing to use the WS-RQAOA with the strongest available relaxations \cite{lasserre2001global, lasserre2006convergent, ghaddar2011second} in the classical pre-processing.
However, higher-order relaxations within these hierarchies  \cite{lasserre2001global,lasserre2006convergent,ghaddar2011second} require a run-time of the classical SDP solver which is super-polynomial in the number $n$ of integral decision variables in \eqref{eqn:qubo} and the order in the hierarchy \cite{lasserre2001global,lasserre2006convergent,ghaddar2011second}. 
Therefore, we limit ourselves to the use of WS-RQAOA with the basic \eqref{eq:SDP} relaxation, whose value can be approximated classically to any fixed precision in polynomial time.

In contrast, one could extend the use of the continuous-valued solution $c^*$ of the  \eqref{eqn:relaxed} relaxation to either the solution $Y^*$ of the basic \eqref{eq:SDP} relaxation, or its strengthened variants \cite{lasserre2006convergent,lasserre2001global}, when preparing the initial state. 
However, this may require more qubits than would be practical in the near-term.
For example, a na\"ive approach to prepare the initial state would utilize $\Theta(n^2)$ and $\Omega(n^2)$ qubits to represent the optimum $Y^*$ of the basic \eqref{eq:SDP} relaxation and its strengthened variants, respectively\footnote{While more elaborate representations of the matrix have been proposed \cite{van2020quantum,brandao2019quantum}, it is not yet clear how to implement the related oracles in practice.}.
At the same time, strong performance guarantees would be readily available for such variants of warm-started quantum optimization. 
For example, consider representing the matrix-valued solution of a \eqref{eq:SDP} relaxation in a $O(n^2)$-qubit initial state, applying a parametrized quantum circuit that allows for the identity in the unitary representation, at least for some choice of its parameters, and then, measuring the qubit register.
This can be seen as a randomized-rounding algorithm, and one can hence analyze the quality of the measured output.

A recently-proposed \cite{abbasi2020sticky} avenue for the analysis of such randomized-rounding algorithms utilizes 
the Sticky Brownian Motion \cite{karlin1981second}, a well-known concept in Stochastic Analysis, possibly with a slowdown due to the use of a speed-function \cite{abbasi2020sticky}, as explained in Appendix \ref{sec:an}. 
In the case of a \eqref{eq:SDP} warm start, 
one can obtain approximation guarantees for rounded solutions that match the best guarantees available classically in polynomial time, see Appendix \ref{sec:an}. 

A particularly important question is whether any of these variants would strictly improve upon the guarantees of GW \cite{Goemans1995}.
Under the Unique Games Conjecture \cite{kempe2007unique,kempe2010unique}, 
it is strictly impossible to improve upon the guarantees of GW using either quantum or classical algorithms unless a quantum computer can solve NP-Hard problems in polynomial time, which is not believed to be the case~\cite{Bennett1997}, or if ${\rm P}={\rm NP}$.
However, a richer picture emerges if this conjecture were to be false and it may be possible to improve approximation ratios using both classical and quantum algorithms.

\section{Simulations with Continuous-Valued Warm-start\label{sec:simulations}}

As a first computational illustration of WS-QAOA, we solve combinatorial-optimization problems framed as a financial-portfolio optimization with a budget constraint \cite{Barkoutsos2020}.
An optimal portfolio minimizes risk and maximizes return by exploiting imperfect correlations in a covariance matrix $\Sigma$ between $n$ assets with expected returns $\mu$ \cite{Markowitz1952}. 
Selecting $B$ assets out of $n$ with equal weights thus requires solving
\begin{align}
    \min_{x\in\{0,1\}^n} qx^T\Sigma x-\mu^T x \\
    \text{such that}\quad \boldsymbol{1}^Tx=B,
\end{align}
where $q$ controls the risk-return trade-off.

\begin{figure}[htbp!]
\centering
\includegraphics[width=\columnwidth]{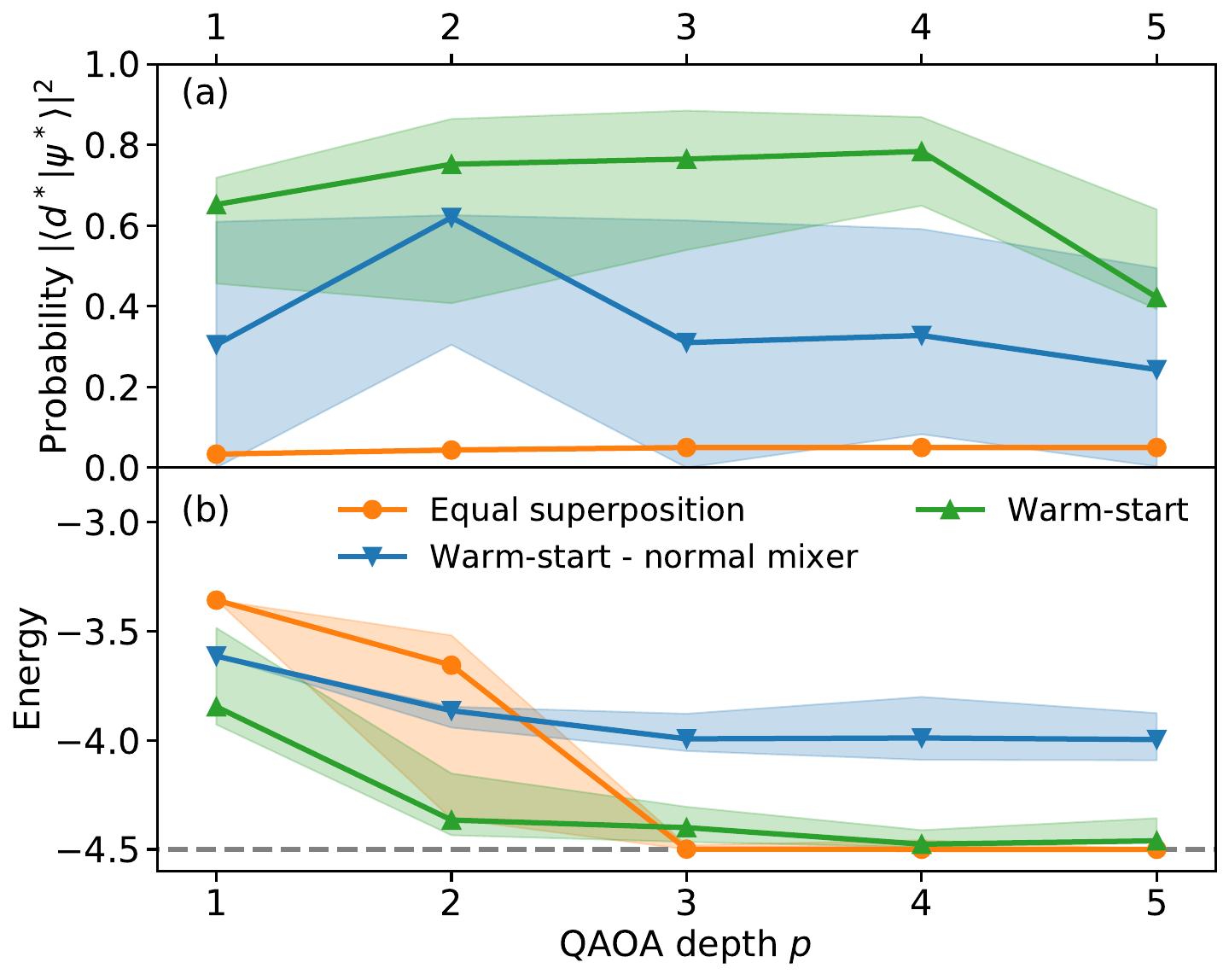}
\caption{(a) Probability to sample the optimal state $\ket{d^*}$ from the optimized trial state $\ket{\psi^*}$ and (b) energy of $\ket{\psi^*}$ for warm-start and standard QAOA at different depths for $n=6$ assets and $q=2$.
The optimal discrete and continuous solutions are $d^*=(0, 0, 1, 1, 1, 0)$ and $c^*\simeq(0.17, 0, 0.97, 0.73, 1.0, 0.14)$, respectively.
QAOA is run ten times with different initial random guesses for $(\boldsymbol{\beta}, \boldsymbol{\gamma})$ chosen uniformly from $\pm 2\pi$.
The thick lines show the median of the ten runs while the shaded areas indicate the 25\% and 75\% quantiles.
The gray dashed line shows $E_0$.
    }
\label{fig:p_evo}
\end{figure}

We create random instances of this problem with $n=6$ assets by simulating the time-series of the asset prices and computing the covariance matrix and returns, see Appendix~\ref{sec:gbm}.
We enforce a budget constraint $B=3$ with a large quadratic penalty term $\lambda(\boldsymbol{1}^Tx-B)^2$ where we chose $\lambda=3$ as it is much larger than $\Sigma$ and $\mu$.
Each instance is mapped to an Ising Hamiltonian $\hat H_C$.
To measure the performance of standard and warm-start QAOA we compute the energy of the optimized trial state $\braket{\psi(\boldsymbol{\beta}^*,\boldsymbol{\gamma}^*)|\hat H_C|\psi(\boldsymbol{\beta}^*,\boldsymbol{\gamma}^*)}$ labeled as $E^*_\text{cold}$ and $E^*_\text{warm}$, respectively.
We normalize $E^*_\text{cold}$ and $E^*_\text{warm}$ to the minimum energy $E_0$ found by diagonalizing $\hat H_C$.
Since the state-vector simulator in Qiskit \cite{qiskit} evaluates the quantum circuits the only source of randomness is the initial guess for $\boldsymbol{\beta}$ and $\boldsymbol{\gamma}$ chosen uniformly from $\pm 2\pi$ given to the COBYLA optimizer we use to find $\boldsymbol{\beta}^*$ and $\boldsymbol{\gamma}^*$.
The optimal solution $c^*$ of the continuous relaxation of the problem used to warm-start QAOA is found with IBM$^\text{\textregistered}$ ILOG$^\text{\textregistered}$ CPLEX$^\text{\textregistered}$ 12.10.0 (CPLEX).
The probability of sampling the optimal binary solution $d^*$ is more than 5 times higher with WS-QAOA then standard QAOA for the simulated depths $1\leq p\leq 5$, see Fig.~\ref{fig:p_evo}(a).
Furthermore, the quality of the solution found by WS-QAOA is better than standard QAOA since $E^*_\text{warm}$ is closer to $E_0$ than $E^*_\text{cold}$, see Fig.~\ref{fig:p_evo}(b).
At depth $p\geq 4$ standard QAOA has enough free parameters to satisfy the budget constraint, as shown by the low energy in Fig.~\ref{fig:p_evo}(b), but still fails to produce a trial state which contains the optimal solution with high probability. 

\begin{figure}[htb!]
    \centering
    \includegraphics[width=\columnwidth, clip, trim=0 10 0 5]{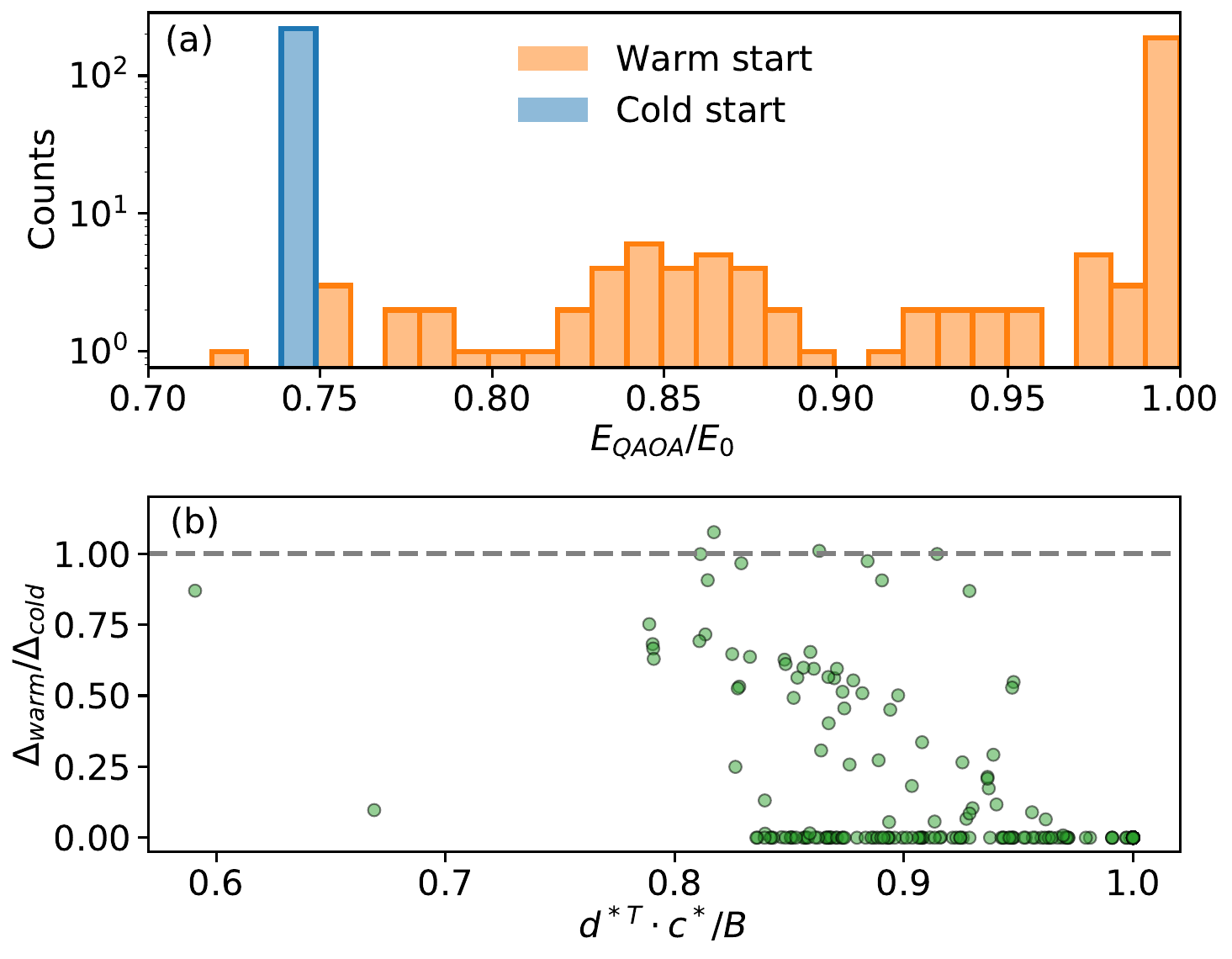}
    \caption{Improvement of depth-one WS-QAOA over standard QAOA for 250 random portfolio instances with $q=2$.
    (a) Histogram of the energy of the optimized trial states $\ket{\psi(\boldsymbol{\beta}^*, \boldsymbol{\gamma}^*)}$ with (orange) and without (blue) warm-start normalized to $E_0$.
    We found $\ket{\psi(\boldsymbol{\beta}^*, \boldsymbol{\gamma}^*)}$ with COBYLA seeded with random initial guesses for $\beta$ and $\gamma$ chosen uniformly from $\pm 2\pi$.
    The minimum energy $E_0$ is found by direct diagonalization.
    (b) Energy difference of WS-QAOA with the optimal solution, i.e. $\Delta_\text{warm}=E^*_\text{warm}-E_0$, normalized to the energy difference obtained with standard QAOA, i.e. $\Delta_\text{cold}=E^*_\text{cold}-E_0$, as a function of the overlap between the optimal solution of the problem with binary weights and continuous weights.
    $\Delta_\text{warm}/\Delta_\text{cold}<1$ implies that WS-QAOA improved the energy of the trial state and $\Delta_\text{warm}/\Delta_\text{cold}=0$ implies that WS-QAOA found the optimal portfolio.
    }
    \label{fig:depth_1}
\end{figure}

We investigate the role of the warm-start mixer operator $\hat H_M^{(ws)}$ by replacing it with the standard mixer $-\sum_{i=0}^{n-1}\hat X_i$ while using the initial state given by the continuous solution $c^*$.
Under these conditions the energy of the optimized state does not converge to the minimum energy, see blue triangles in Fig.~\ref{fig:p_evo}(b).
The probability of sampling the optimal discrete solution is between warm-start and standard QAOA but depends heavily on the initial point given to COBYLA, see Fig.~\ref{fig:p_evo}(a).
These results further justify the use of the modified mixer in WS-QAOA.

To further illustrate the advantage of a warm-start at low depth we solve 250 random portfolio instances with warm-start and standard QAOA, both at depth $p=1$.
Here, the standard QAOA produces variational states that poorly approximate the ground state, see the histogram of $E^*_\text{cold}$ in Fig.~\ref{fig:depth_1}(a).
However, WS-QAOA produces optimized variational states that are much closer to the minimum energy of each problem Hamiltonian.
Furthermore, we find that WS-QAOA tends to produce better solutions when the overlap $d^{*T}c^*/B$ between the optimal solutions to the discrete and relaxed problems is closer to 1, see Fig.~\ref{fig:depth_1}(b).

\begin{figure}[tb!]
    \centering
    \includegraphics[width=\columnwidth]{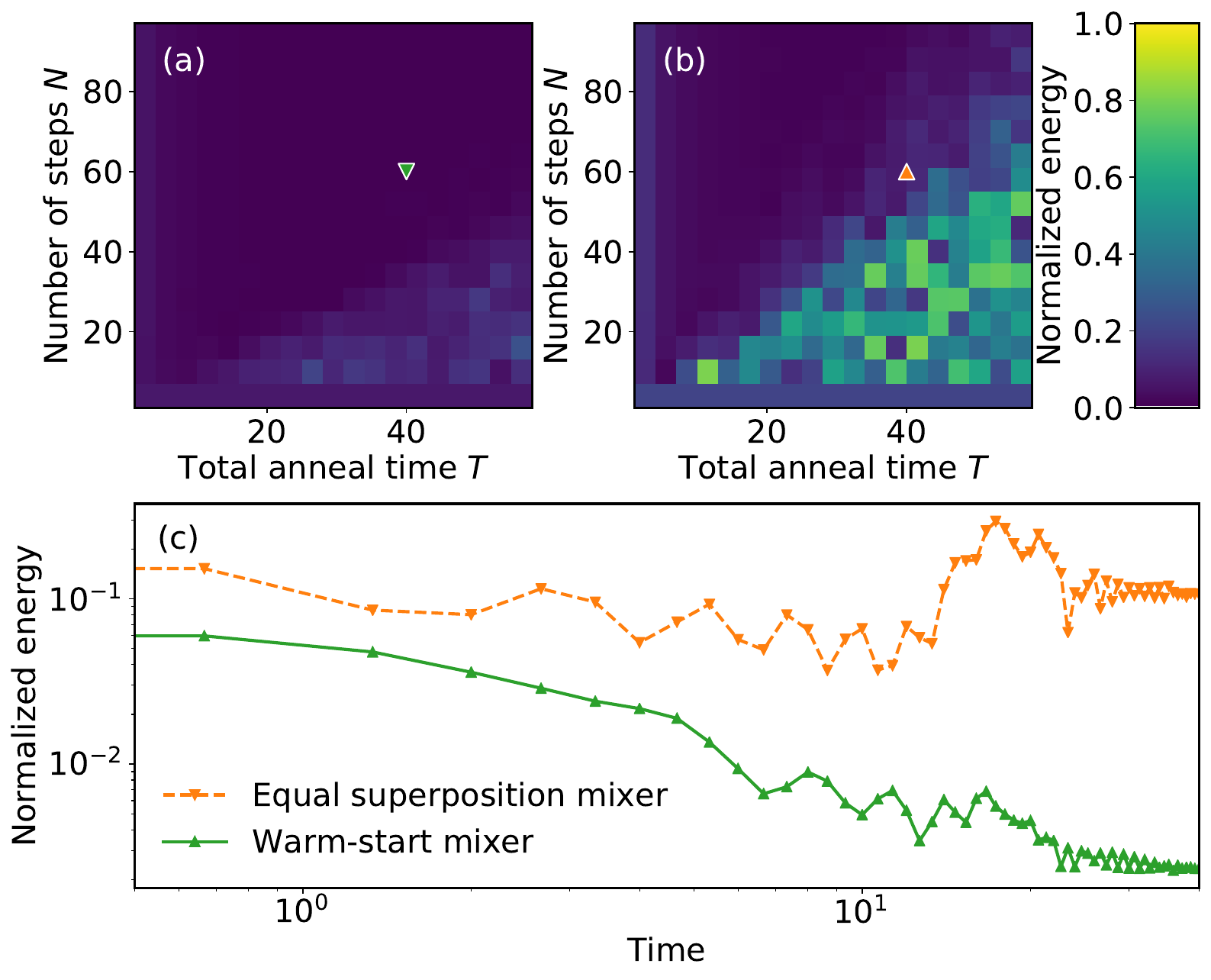}
    \caption{Energy of the quantum state normalized to $[\min(\hat H_{C}), \max(\hat H_{C})]$ after Trotterized annealing for the warm-start mixer (a) and the equal superposition mixer (b). The triangles in (a) and (b) indicate the points for which the time-resolved energy in (c) was obtained.}
    \label{fig:portfolio_anneal}
\end{figure}

We now illustrate the advantage of a warm-start in the context of quantum annealing.
We simulate annealing by Trotterizing the time-evolution starting form the ground state of the mixer Hamiltonian.
At time step $k\in\{0,..., N\}$ we apply the operator
$\exp(-i\beta_k\hat H_\text{M})\exp(-i\gamma_k\hat H_\text{C})$ where the annealing schedules are $\beta_k=2\delta t(1-i/N)$ and $\gamma_k=2i\delta_t/N$ with $\delta t=T/N$.
We compute the energy as function of $T$, which controls the total annealing time, and $N$ using the equal-superposition mixer $\sum_{i=0}^{n-1}\hat X_i$ and the warm-start mixer $\hat H_{M}^{(ws)}$.
Warm start requires less annealing time $T$ and a smaller number of steps $N$ than the equal superposition mixer to minimize the energy,  as can be seen by comparing (a)~and~(b) in Fig.~\ref{fig:portfolio_anneal}.
This is further confirmed by computing the energy at each time step for $T=40$ and $N=60$.
When the warm-start mixer $\hat H_{M}^{(ws)}$ is used, the initial and final energies are lower than when the equal superposition mixer is used.
These data emphasize, from an annealing point of view, that WS-QAOA converges faster than QAOA.

\section{Simulations with Rounded Warm-Start\label{sec:max_cut}}

Next, we discuss warm-starting QAOA for MAXCUT.
The maximum cut of an edge-weighted graph $G=(V, E)$ with nodes $V$, edges $E$, and weights $\omega_{ij}, \{i, j\} \in E$ is a partitioning of the set of nodes $V$ in two such that the sum of the edge weights $\omega_{ij}$ where $i$ and $j$ are in different parts is maximized.
This is cast as
\begin{align}\label{eq:maxcut}
    \max \frac{1}{2}\sum_{(i,j) \in E}\omega_{ij}(1-z_iz_j) \\
    \text{such that}\quad z\in\{-1,1\}^{|V|}, \notag
\end{align}
where the binary variable $z_i$ indicates which side of the cut node $i$ is on.
In the case of positive edge weights $\omega_{ij}$, for any $\epsilon$, the problem cannot be approximated within the ratio of  $16/17 - \epsilon$ classically in polynomial time \cite{Hastad2001}, unless P = NP. 
In the case of the real-valued edge-weights $\omega_{ij}$, the hardness factor is $11/13 - \epsilon$ \cite{charikar2004maximizing}.
In both cases, under the Unique Games Conjecture \cite{khot2007optimal,KhotSurvey,khot2015unique}, the best guarantees obtainable classically in polynomial time are those of the random-hyperplane rounding \cite{Goemans1995,Karloff1999,charikar2004maximizing}, as we detail in Appendices~\ref{sec:gw} and \ref{sec:gw2}.

\begin{figure}[tb!]
    \centering
    \includegraphics[width=\columnwidth, clip, trim=0 10 0 10]{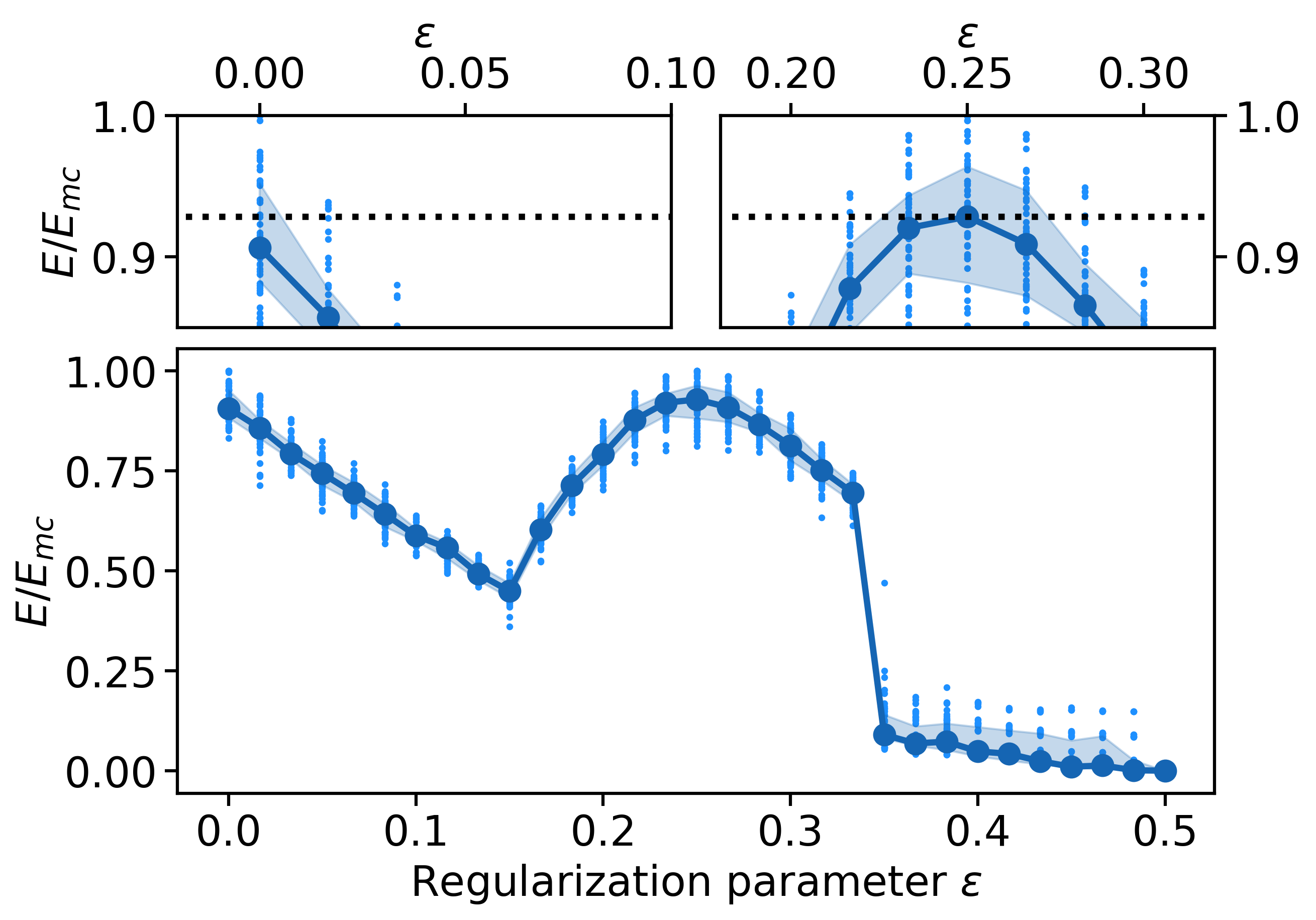}
    \caption{Energy, normalized to the energy of the maximum cut, as a function of $\varepsilon$ for ten graphs each solved five times with different GW cuts. 
    The shaded area indicates the 25\% to 75\% quantiles and the line shows the median.
    Each small dot is the energy from one WS-QAOA.
    The dotted line shows the median normalized energy at $\varepsilon=0.25$.}
    \label{fig:epsilon_graph}
\end{figure}

Since we cannot relax MAXCUT to a (\ref{eqn:relaxed}) with a positive semi-definite matrix, without a projection
onto the cone of positive-semidefinite matrices, we now warm-start QAOA with a \emph{binary} solution obtained using the GW rounding.
Here, the variational form can only produce states different from the initial GW cut when the regularization parameter $\varepsilon>0$.
We study the effect of $\varepsilon$ by minimizing the energy of depth-one WS-QAOAs applied to ten fully connected graphs with 30 nodes and edge weights uniformly chosen from $\{-10,-9,...,0, ..., 9, 10\}$.
For each graph we generate ten GW cuts and study the five best cuts with WS-QAOA.
To find the optimal $\beta_1$ and $\gamma_1$ we seed COBYLA with an initial point obtained as follows.
We perform a grid search in $\gamma_1$ with 25 equally spaced values from $-2\pi$ to $2\pi$.
For each value of $\gamma_1$ we evaluate the energy for the two points $\beta_1\in\{3\pi/4, 3\pi/8\}$ and fit the result to $a\sin(2\beta_1)+b\sin(4\beta_1)$ to find the $\beta_1$ that minimizes the energy without having to perform a two-dimensional grid scan.
The $(\beta_1,\gamma_1)$ with the lowest energy is given as initial point to COBYLA.
At $\varepsilon=0$ the median energy, normalized to the energy of the maximum cut, is 0.907 and corresponds to the energy of the GW cuts used to warm-start QAOA, see Fig.~\ref{fig:epsilon_graph}.
As $\varepsilon$ increases, the normalized energy decreases.
However, around $\varepsilon=0.15$ the median energy starts to increase, and for $\varepsilon=0.25$ rises beyond the energy of the GW cut to 0.929, which suggests that warm-starting quantum optimization may lead to algorithms which can outperform the GW randomized rounding.
This increase in $\varepsilon$ is not observed when the mixer from Eq.~\eqref{eq:new_mixer} is used, see Appendix \ref{sec:ws_mixer_mc}.

Next, we illustrate the WS-RQAOA algorithm outlined in Sec.~\ref{sec:rounded} at depth one by searching for the maximum cut of arbitrary graphs with $n=20$, and $n=30$ nodes.
Two types of graphs are solved, one where each edge appears with a $p_E=1/2$ probability and has a $1/2$ probability of having a positive or negative unit weight.
The second type of graphs are fully connected with uniformly distributed edge weights sampled from $\{-10,-9,...,0, ..., 9, 10\}$.
We expect that finding the maximum cut for the fully connected graphs will be harder than those with $p_E=1/2$ \cite{Akshay2020b} and that the resulting QAOA circuits will be deeper as they have more edges \cite{Ostroswki2020}.
For each graph size and type we randomly generate 100 graphs. 
At each iteration a GW pre-solver generates $N=10$ cuts of which we select the best $M=5$ unique cuts to warm-start five QAOA solvers with a depth $p=1$ and $\varepsilon=0.25$, chosen based on the results from Fig.~\ref{fig:epsilon_graph}.
We chose a low $N$ to avoid systematically giving QAOA the maximum cut, see Appendix \ref{sec:gw}.
This only holds for the small graphs with which we illustrate WS-RQAOA. 
For larger graphs we would, however, choose a much larger $N$ as GW cuts can be efficiently generated.
The standard and warm-start depth-one RQAOA algorithms are efficiently simulated by computing the correlations 
$\langle \hat Z_i\hat Z_j\rangle$ 
at each iteration, see Appendix \ref{sec:efficient_rqaoa}.
The parameters $\beta_1$ and $\gamma_1$ are optimized with COBYLA which is initialized with a good initial point obtained from a grid search, as discussed above, to avoid local-minima.
When a graph reaches $n_\text{stop}=n/2$ nodes we diagonalize the corresponding Hamiltonian to find the maximum cut of this reduced problem.
Together with the replacements from RQAOA we obtain an approximation of the maximum cut of the original graph with $n$ nodes.
We compare WS-RQAOA with standard RQAOA.

\begin{figure*}[htb!]
    \centering
    \includegraphics[width=\textwidth, clip, trim=0 10 0 10]{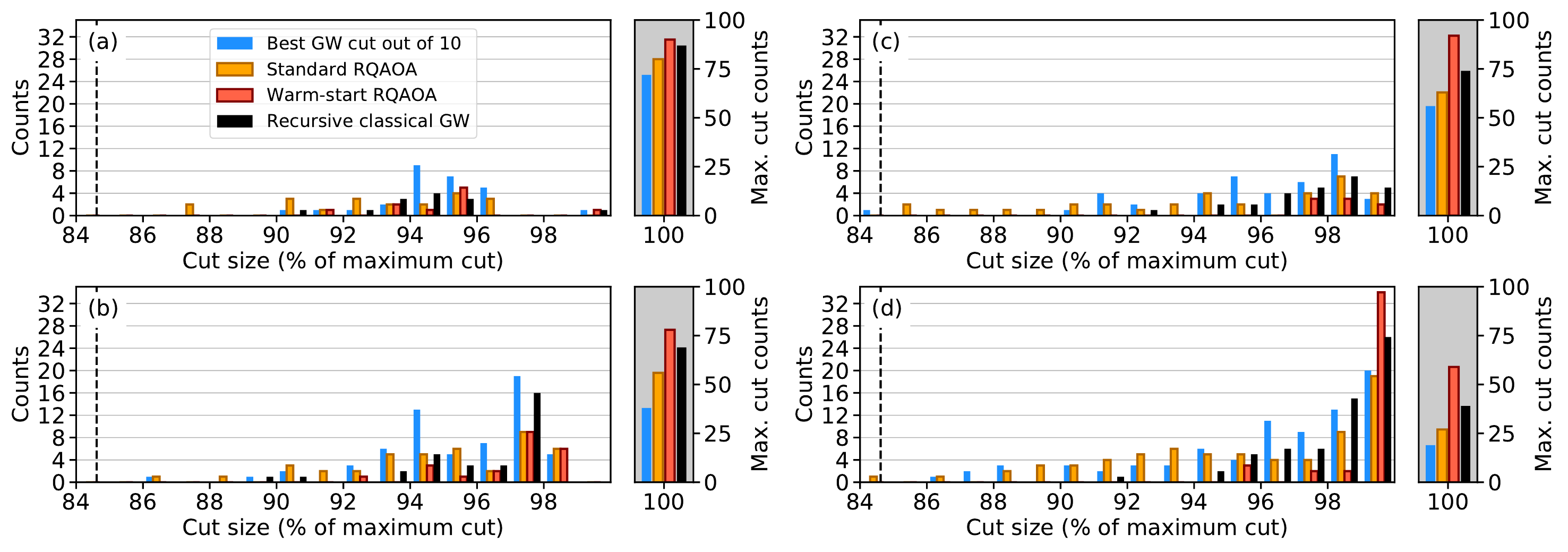}
    \caption{Histograms of cut sizes, relative to the maximum cut found by CPLEX, for the best out of ten cuts generated by GW on the initial graph (blue), standard RQAOA (orange), WS-RQAOA (red), and the recursive classical solver based on GW (black).
    (a) and (b) correspond to random graphs with 20 and 30 nodes, respectively, $p_E=1/2$ and edge weights in $\{-1,1\}$.
    (c) and (d) correspond to fully-connected graphs with 20 and 30 nodes, respectively, with edge weights uniformly distributed in $\{-10,-9, ..., 0, ..., 10\}$.
    The number of maximum cuts found is shown in the gray shaded sub-plots and not the main plot. 
    The dashed line shows the hardness factor $11/13$.
    }
    \label{fig:max_cut}
\end{figure*}

\begin{figure}[htb!]
    \centering
    \includegraphics[width=\columnwidth, clip, trim=0 3 0 5]{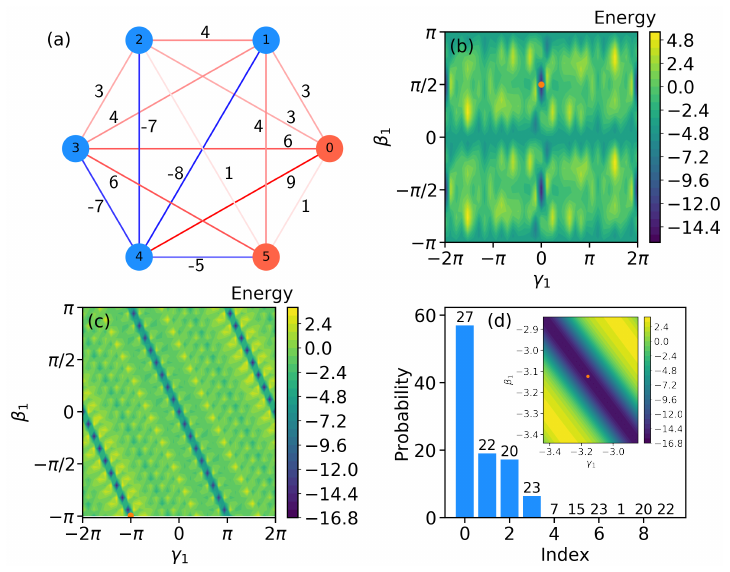}
    \caption{(a) The graph used to study the energy-landscape as a function of $\beta_1$ and $\gamma_1$ for depth-one (b) and depth-three (c) WS-QAOA.
    The initial cut is 001111 and $\varepsilon=0.25$.
    Edge and node colors indicate the edge weight and maximum cut with value 27, respectively.
    (d) Ten highest probability cuts in the optimized depth-three trial state.
    The numbers indicate the cut-size.
    In (c) the values of $\beta_i$ and $\gamma_i$ for $i=2,3$ are given by COBYLA after minimizing the energy $E(\boldsymbol{\beta},\boldsymbol{\gamma})$ and correspond to the best point in Fig.~\ref{fig:depth_study_mc}.
    The inset in (d) is a zoom of (c) around the optimal point (orange dot) found by COBYLA.}
    \label{fig:graph_example}
\end{figure}

Our simulations indicate that WS-RQAOA outperforms standard RQAOA, see Fig.~\ref{fig:max_cut}, and that the number of maximum cuts found decreases with graph size.
Fully connected graphs with 30 nodes are the hardest to solve among the graphs we consider.
Still, for the graphs in Fig.~\ref{fig:max_cut}, we observe that the optimal $\beta^*_1$ and $\gamma^*_1$ are systematically found.
This indicates that when warm-start and standard RQAOA fail, it is because the depth-one variational form 
is not versatile enough to capture the correlations in the maximum cut.
At $\varepsilon=0.25$ we often observed that $\beta^*_1=\pi/2$ and $\gamma^*_1=0$, see Fig.~\ref{fig:graph_example}.
We therefore benchmark WS-RQAOA against a classical recursive optimization procedure,  where the average correlation matrix of the five best GW cuts is used to eliminate   decision variables in each iteration,
similarly to RQAOA, 
see black bars in Fig.~\ref{fig:max_cut}.
This classical algorithm performs better than standard RQAOA, but slightly worse than WS-RQAOA.

\begin{figure}[htbp!]
    \centering
    \includegraphics[width=\columnwidth, clip, trim=0 10 0 5]{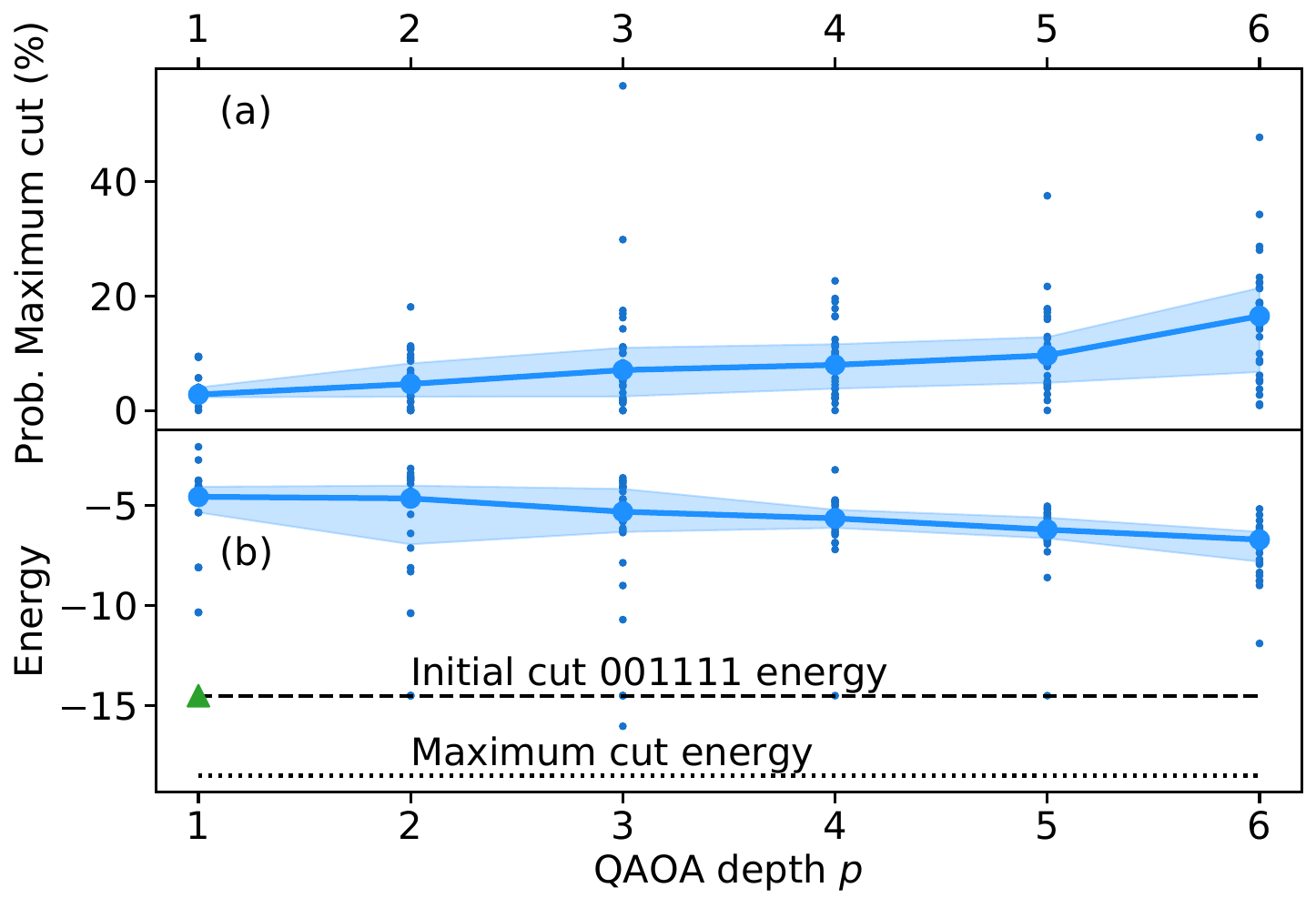}
    \caption{(a) Probability of sampling the maximum cut of the six-node graph shown in Fig.~\ref{fig:graph_example}(a) from the optimized trial state and (b) its energy as a function of QAOA depth $p$.
    The shaded areas indicate the 25\% and 75\% quantiles of 30 runs represented as small dots.
    Their medians are the large dots.
    The green triangle is the energy of the depth-one trial state with $\beta_1=\pi/2$ and $\gamma_1=0$.}
    \label{fig:depth_study_mc}
\end{figure}

We now investigate WS-QAOA for $p>1$ in a non-recursive setting.
Since the efficient algorithm outlined in Appendix~\ref{sec:efficient_rqaoa} is not valid for $p>1$ we solve a small, fully connected graph with six nodes and edge weights uniformly distributed in $\{-10, -9, ..., 0, ..., 10\}$, see Fig.~\ref{fig:graph_example}(a).
The maximum cut of this graph has size 27.
By comparing the energy landscape $E(\beta_1, \gamma_1)$ of a depth-one and depth-three WS-QAOA initialized with the cut 001111, of size 23, we observe that the optimal trial state of deeper variational forms is no-longer the initial GW cut, see Fig.~\ref{fig:graph_example}(b-d).
We study the probability of sampling the maximum cut as a function of $p$ by running 30 WS-QAOAs each with a random initial $\boldsymbol{\beta}$ and $\boldsymbol{\gamma}$ chosen uniformly from $[0, 2\pi]$ and $\pm2\pi$, respectively, for depths $p=1,...,6$.
The probability to sample the maximum cut for this graph increases with $p$ while the energy of the optimized trial state decreases, see Fig.~\ref{fig:depth_study_mc}.
This matches our expectations from theory since the circuit at depth $p+1$ can reproduce all states of the depth $p$ variational form while being more flexible.
Since the energy landscape is non-convex and contains many local minima it is challenging to find globally optimal parameters starting from random guesses of $\boldsymbol{\beta}$ and $\boldsymbol{\gamma}$ \cite{Wang2018, Shaydulin2019, Zhou2020, Larkin2020}.
Even at depth-one with random initial guesses for $\beta_1$ and $\gamma_1$ COBYLA does not always find the optimal $\beta_1=\pi/2$ and $\gamma_1=0$, see Fig.~\ref{fig:graph_example} and Fig.~\ref{fig:depth_study_mc}(b).
The complexity of the energy landscape, even for this six-node graph, may therefore explain why the energy of the optimized trial state decreases slowly with $p$.

\section{Discussion and Conclusion\label{sec:conclusion}}

We hope to have contributed towards a framework for the design of quantum optimization algorithms with a warm start, and towards reasoning about their properties.
Currently, these algorithms can achieve the same guarantees as the classical relaxations upon which they are based.
If the Unique Games Conjecture is true, these guarantees cannot be improved upon by classical or quantum algorithms running in polynomial time, unless we can solve NP-Hard problems in polynomial time.
However, if this conjecture is false then both quantum and classical algorithms may be able to improve the existing performance guarantees.

An implementation of WS-QAOA is available in Qiskit~\cite{QiskitOptim}, 
the open-source software development kit 
for working with quantum computers.
Our simulations show that warm-starting quantum heuristics provides an advantage at low depth.
This is particularly important for dense optimization problems intended to be solved on noisy quantum hardware that struggles to implement deep quantum circuits.
The portfolio optimization simulations indicate that WS-QAOA finds better solutions than standard QAOA.
Here, future work could investigate tying budget constraints into the quantum circuit of WS-QAOA \cite{Bartschi2020}.

We have also demonstrated how to continuously transform WS-QAOA to conventional QAOA using the regularization parameter $\varepsilon$.
We also showed how to improve QAOA for MAXCUT using the GW algorithm to warm-start RQAOA, albeit by introducing an inconsistency between the mixer and the initial state.
By using a grid scan at depth one, we mitigated the effect of local optima.
We applied WS-RQAOA at depths $p>1$ on a single graph.
The results suggest that the algorithm provides better solutions as $p$ increases.
Future work may extend these simulations to more graphs with different sizes.
Further work could also exploit other possible warm-starts, e.g., based on polynomially-solvable special cases  \cite{allemand2001polynomial,hladik2019new}, where one could for example consider low-rank approximations of~$\Sigma$ or the SDP \cite{Tate2020} as well as analysis of the convergence properties when using a modified mixer that does not have the initial state as eigenstate.
One may also investigate RQAOA in the context of the continuous warm-start discussed in Sec.~\ref{sec:simulations}.
Furthermore, we may also warm-start quantum optimization algorithms from candidate solutions obtained with classical solvers such as CPLEX or GUROBI with a time-limit termination criterion.

We expect warm-start to be applicable to other problems within Combinatorial Optimization and Integer Programming, for which a good solution can be found through randomized rounding \cite{Raghavan1987},
possibly following an encoding into a QUBO \cite{Lucas2014,dunning2018works,lodewijks2019mapping},
a mixed-integer linear optimization problem \cite{LASSERRE2016},
or a polynomial unconstrained binary optimization problem \cite{Ostroswki2020}.
Indeed, both the recipe to obtain SDP relaxations \cite{poljak1995recipe} and the analytical tools of Appendix \ref{sec:an} are applicable to linearly constrained problems equally well. 
For example, the particle-hole representation for VQE can be seen as a form of warm-start \cite{Barkoutsos2018}.
We anticipate that WS-QAOA is also applicable to other binary optimization problems for which an approximate solution can easily be found using relaxed versions of the problem, without the use of randomized rounding, albeit more research needs to be done in this direction.

\section*{Acknowledgements}
The authors acknowledge useful discussions with Libor Caha, Donny Greenberg, Sergey Bravyi and Giacomo Nannicini.
IBM, the IBM logo, and ibm.com are trademarks of International Business Machines Corp., registered in many jurisdictions worldwide. Other product and service names might be trademarks of IBM or other companies. The current list of IBM trademarks is available at https://www.ibm.com/legal/copytrade.
Jakub Mare\v{c}ek's research has been supported by the OP RDE project CZ.02.1.01/0.0/0.0/16\_019/0000765 Research Center for Informatics.
One month after releasing our work Ref.~\cite{Tate2020} appeared on the arXiv.

\FloatBarrier
\appendix

\section{The Unique Games Conjecture\label{sec:ucg}}

We now summarize the Unique Games Conjecture (UGC) without presenting any original material.
Inapproximability results suggest that finding an approximate solution of a certain problem-dependent approximation ratio is no easier than finding the optimal solution. 
The PCP theorem \cite{arora1998probabilistic,arora1998proof,Dinur2007} shows
that in a constraint satisfaction problem, the fraction of satisfiable constraints is NP-Hard to approximate within \emph{some} constant factor.
In particular, for a constraint satisfaction problem with at most $k$ variables per constraint, it suggests
there is a constant $0 < \alpha < 1$, such that it is NP-Hard to decide whether either all constraints are simultaneously satisfiable or whether every assignment satisfies fewer than an $\alpha$ fraction of the constraints.
We refer to \cite[Chapters 18 and 19]{arora2009computational} for an excellent overview. 

\begin{figure}[htb!]
    \centering
    \includegraphics[width=0.9\columnwidth, clip, trim=13 40 390 20]{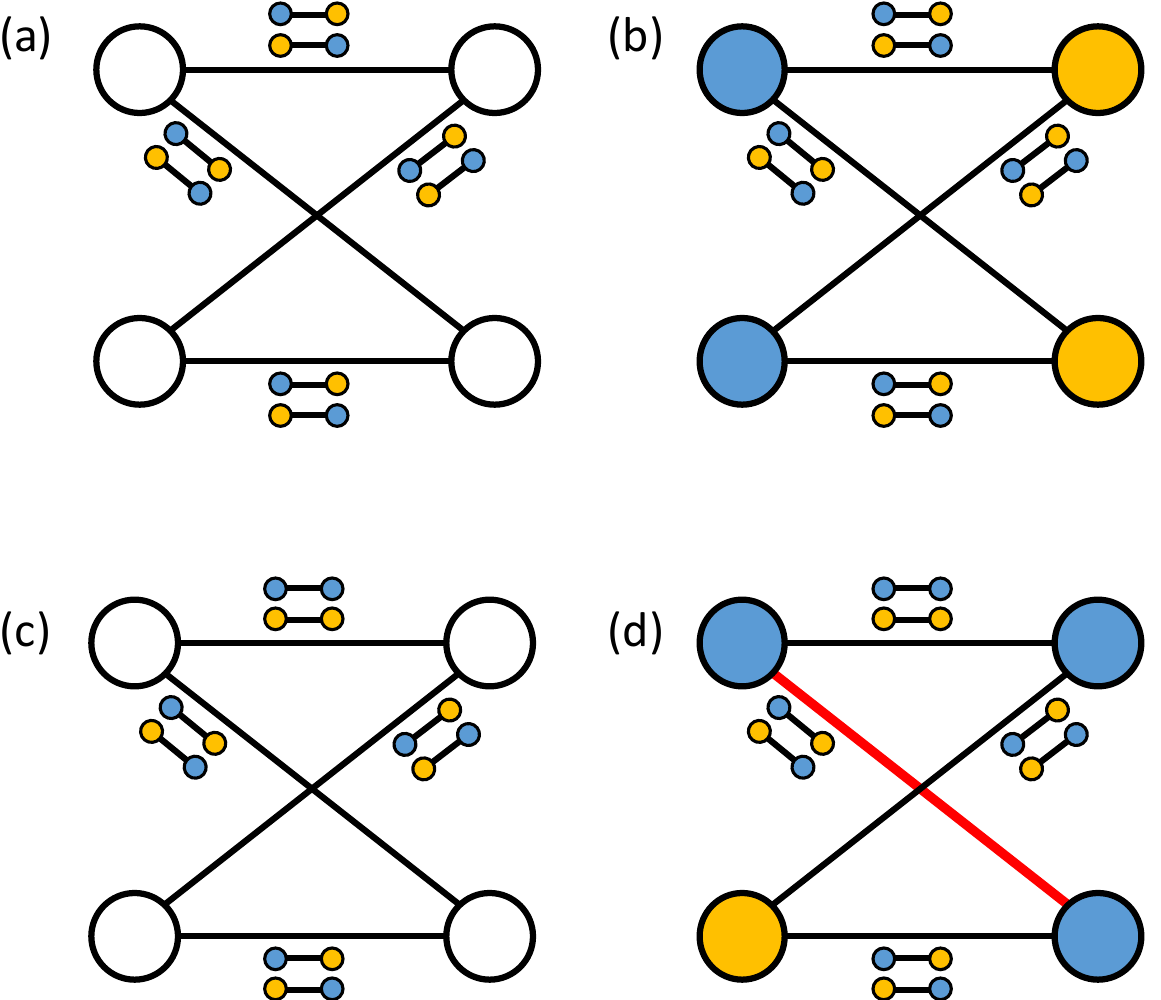}
    \caption{Unique label cover. (a) and (c) are examples of unique label cover problems with a graph with four nodes and four edges denoted $\mathcal{L}_a$ and $\mathcal{L}_c$, respectively.
    The alphabet has two colors: orange and blue.
    (b) An assignment of node colors for $\mathcal{L}_a$ where all edges are satisfied, i.e. $\opt(\mathcal{L}_a)=1$.
    (d) An assignment of node colors for $\mathcal{L}_c$ where one edge is not satisfied. 
    We have $\opt(\mathcal{L}_c)=3/4$.}
    \label{fig:ulc_example}
\end{figure}

By building on the PCP theorem Khot suggested the Unique Games Conjecture~\cite{khot2002power} which can be formulated with two-prover one-round games as well as unique label cover problems.
Here, we state the UGC with the unique label cover problem as it relates to MAXCUT~\cite{khot2007optimal}.
In a unique label cover problem there is a bipartite graph $G=(V,W,E)$ with partition $V$, $W$ and edges $E$, an alphabet $M$, and a bijection $\pi_e \colon M \rightarrow M$ for every edge $e \in E$. 
We denote this by ${\mathcal L}(V,W,E,M,\{\pi_{e}\}_E)$.
Given the assignments ${\mathcal A}_V:V\to M$ and ${\mathcal A}_W:W\to M$,
an edge $e = (v, w) \in E$ is satisfied if and only if $\pi_e({\mathcal A}_W(w))={\mathcal A}_V(v)$. 
Now let $\opt(\mathcal{L})$ be the maximum weight of edges satisfied by any assignment.
The Unique Label Cover problem with parameter $\delta$, denoted ULC$(\delta)$, is the problem of deciding, given an instance ${\mathcal L}(V,W,E,M,\{\pi_{e}\}_E)$, whether 
$\opt({\mathcal L}) \geq 1-\delta$ or $\opt({\mathcal L}) \leq \delta$, see examples in Fig.~\ref{fig:ulc_example}. Then:

\begin{conjecture}[Unique Games Conjecture \cite{khot2002power}]
For arbitrarily small constants $\zeta,\delta>0$, there exists a constant $k=k(\zeta,\delta)$ such that it is NP-Hard to determine whether a unique label cover instance with the label set size $k$, i.e. $k=|M|$, has optimum at least $1-\zeta$ or at most $\delta$.
\end{conjecture}

Under the UGC a canonical semidefinite programming relaxation of a constraint satisfaction problem provides the best possible approximation ratio \cite{khot2007optimal,raghavendra2008optimal,raghavendra2009round}.
Nevertheless, the UGC itself has been neither proven nor disproven, despite much effort.
The closest to proving the UGC, namely Khot \emph{et al.}~\cite{Khot2018}, essentially proved the so-called 2-to-2-Games Conjecture, which is a specialisation of the UGC to finite fields on two elements.
Independently, \cite{barak2011rounding,hopkins2020subexponential} have shown that there is a
subexponential time approximation for the original Unique Games problem,
which utilizes a hierarchy of LP \cite{hopkins2020subexponential} or SDP \cite{barak2011rounding} relaxations. 
This seems to point to the full UGC being true, but does not prove it yet.

Research in quantum complexity is still ongoing.
The EPR paradox \cite{einstein1935can} and the Tsirelson problem \cite{cirel1980quantum} can be seen as two-prover one-round games in which the provers may share entanglement while the verifier and all communication are classical. 
In this spirit, Kempe \emph{et al.}~\cite{kempe2010unique} showed that the ``unique games with entangled provers'' is false by rounding an SDP relaxation of the value of a unique game with entangled provers and with more than two possible answers, in polynomial time.
The ``games quantum PCP conjecture'', where one wishes to distinguish between the cases when the provers of a game with a classical verifier have a strategy using entanglement that succeeds with probability 1, or when no such strategy succeeds with probability larger than 1/2, has been shown to be true  under randomized reductions~\cite{8555153}.
By contrast, the ``constraint satisfaction'' variant of the quantum PCP conjecture \cite{aharonov2009detectability}, which considers constant-factor approximations to the minimal energy of a local Hamiltonian normalized to 1, remains open.

\section{Goemans-Williamson Algorithm\label{sec:gw}}

The GW algorithm \cite{Goemans1995} first solves the continuous relaxation of MAXCUT
\begin{align} \label{eqn:mc_relaxed} 
    \max\frac{1}{2}\sum_{i<j}\omega_{ij}(1-v_i^T v_j) 
\end{align}
with positive edge weights $\omega_{ij}$, 
where the decision variables $v_i$ are $n$-dimensional vectors with unit Euclidean norm instead of binary variables $z_i\in\{-1,1\}$.
We denote this $v_i \in S^{n}$ with $S$ in plain font, in contrast to $\mathbb{S}^{n}$ for symmetric matrices.
The relaxation~(\ref{eqn:mc_relaxed}) is efficiently solvable as a semidefinite programming problem \cite{anjos2011handbook} to get the optimal vectors $v_i^*$.

Next, the GW algorithm generates a cut by selecting a vector $r$ uniformly at random on the unit sphere and 
assigning $z_i=\sgn (r^Tv_i^*)$ for each node, where the $\sgn$ function returns $1$ for non-negative inputs and -1 elsewhere. 
That is, the rounding depends on which side of the hyperplane (defined by $r$) passing through the origin the node lies.

Informally speaking, cuts generated in this way are guaranteed to be on average $87.9\%$ of the size of the maximum cut \cite{Goemans1995}, when averaging over the choice of the random hyperplane in the case of the positive edge weights. Formally, 
\begin{proposition}[Based on Theorem 3.1 in \cite{Goemans1995}]
The expected value, with respect to the random hyper-plane defined by the vector $r$,
of the cut size $W$ generated by rounding of the MAXCUT SDP relaxation \eqref{eqn:mc_relaxed} is: 
\begin{align}
 \mathbb{E} \big[W \big]\; = & \sum_{1 \le i<j\le n} \omega_{ij } \Prob[\sgn(r^T v_i) \not = \sgn(r^T v_j)] \notag \\
 = \; &
 \frac{1}{\pi} \sum_{1 \le i<j\le n} \omega_{ij }\arccos(v_i^T v_j) \notag \\
 \ge \; & \alpha W^*,
\end{align}
where $W^*$ denotes the value of the maximum cut and the hardness factor is
\begin{align}
\alpha ={\frac {2}{\pi }}\min _{0\leq \theta \leq \pi }{\frac {\theta }{1-\cos \theta }} \approx 0.878.
\label{eq:alphadef}
\end{align}
\label{prop:0}
\end{proposition}

Further, conditional on the Unique Games Conjecture \cite{khot2007optimal,KhotSurvey,khot2015unique}, this is the best possible guarantee that can be obtained by any classical algorithm in polynomial time. 

\begin{figure}[htb!]
    \centering
    \includegraphics[width=\columnwidth, clip, trim=0 10 0 10]{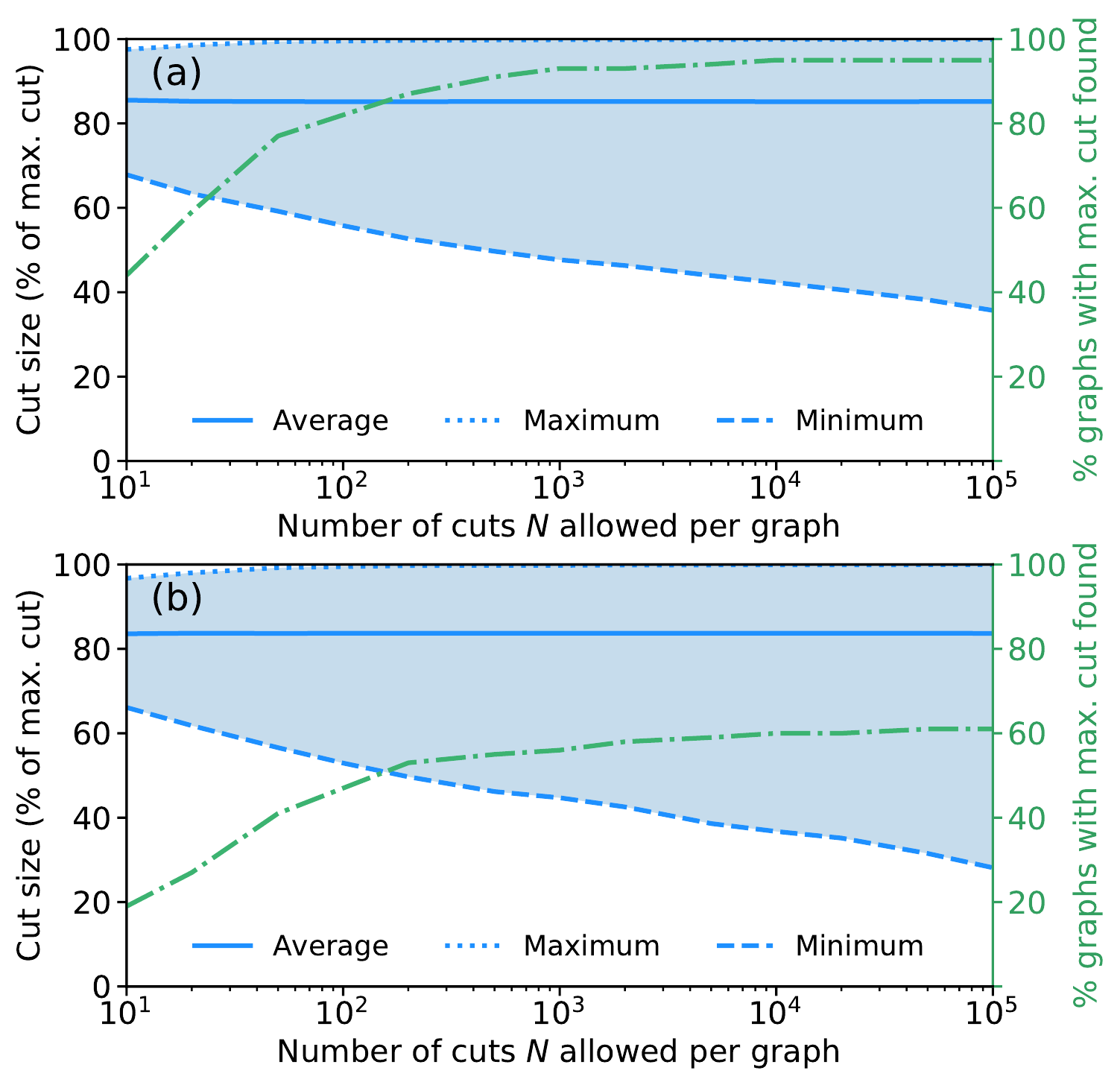}
    \caption{(a) and (b) show the size of GW cuts relative to the maximum cut for the graphs in Fig.~\ref{fig:max_cut}(b) and (d), respectively.
    For each graph, $N$ GW cuts are generated and the averages, minima, and maxima of the per-graph objective-function values are computed.
    These averages, minimums, and maximums are then averaged over the 100 graphs.
    The dotted-dashed green line shows the fraction of graphs for which the maximum cut was found.
    It is harder to find the maximum cut on fully connected uniform random graphs than random graphs with $p_E=1/2$.
    }
    \label{fig:gw_100_graphs}
\end{figure}

\section{Extensions towards QUBO \label{sec:gw2}}

MAXCUT is a special case of \eqref{eqn:qubo}\footnote{Consider the matrix $\Sigma$, where for each edge weight $\omega_{ij}$ there is an entry $\Sigma_{ij} = \Sigma_{ji} = - \omega_{ij}$, and for all other values there are zeros.}.
The GW performance ratio is valid only for MAXCUT, as the special case of \eqref{eqn:qubo} \cite{charikar2004maximizing}, and likewise the constants in the inapproximability results.

One can, however, encode most problems in combinatorial optimization into a so-called constraint satisfaction problem \cite{charikar2004maximizing,charikar2006near,raghavendra2008optimal}, for which there is an well-known SDP relaxation and a subsequent rounding procedure \cite{charikar2004maximizing,charikar2006near,raghavendra2008optimal}. Likewise, one can derive optimal inapproximability results  conditional on the Unique Games Conjecture \cite{khot2007optimal,KhotSurvey,khot2015unique}. See, for example, Figure 2 of \cite{KhotSurvey}.

For example, for MAXCUT with real-valued edge weights \cite{charikar2004maximizing}, which actually generalises the QUBO we have presented, as it does not assume $\Sigma$ is symmetric, we have:

\begin{proposition}[Based on Lemma 6 in \cite{charikar2004maximizing}]
Let $w$ stand for
the total weight of edges in a MAXCUT instance,
where it is NP-Hard to decide whether the optimal cut is larger or equal than $k$ or less than $\alpha k$, where $\alpha$ is the hardness factor  \eqref{eq:alphadef} for MAXCUT.
Then for every $\epsilon > 0$ it is NP-Hard to distinguish instances of QUBO with optimum greater or equal to $2k - w$ from instances of QUBO whose optimum is at most $2 \alpha k - w$.
The ratio of these two bounds on the optimum is
\begin{align}
\frac{2 \alpha k - w}%
{2k - w} = \alpha + w \frac{ \alpha - 1}{2k - w}.
\end{align}
\label{prop:QUBO}
Moreover, the optimum hardness factor is achieved by the randomized rounding of an SDP relaxation \cite{charikar2004maximizing}.
\end{proposition}

This can be used to prove the inapproximability results for MAXCUT with real weights \cite{charikar2004maximizing}, both conditional and independent of the Unique Games Conjecture.

We illustrate the performance of GW on the random graphs with 30 nodes used in Sec.~\ref{sec:max_cut}.
For each graph we generate $N$ cuts with GW and normalize them to the maximum cut which is found with CPLEX.
We next calculate the minimum, maximum, and average size of these $N$ cuts for each graph.
Finally, we average the minimum, maximum, and average of the 100 graphs, see Fig.~\ref{fig:gw_100_graphs}.
The average is stable at $85.2\%$ and $83.7\%$ for the random graphs with $p_E=1/2$ and the fully connected graphs, respectively, see Fig.~\ref{fig:gw_100_graphs}(a) and (b).
These averages are slightly below the GW approximation ratio\footnote{This could be seen as an instance of a phase transition \cite{achlioptas2005rigorous}, beyond which the problem becomes computationally difficult for classical algorithms running in polynomial time. Notice that the MAX-2-SAT of \cite{achlioptas2005rigorous} is closely related to MAXCUT \cite{coppersmith2004random}.}.
When $N>100$ the maximum cut for more than 80 of the 100 graphs in Fig.~\ref{fig:gw_100_graphs}(a) is found which is why we chose $N=10$ in Sec.~\ref{sec:max_cut}.
When the graphs are fully connected the GW algorithm does not find as many maximum cuts.
For instance, 61 maximum cuts are found at $N=10^5$ for fully connected graphs, see Fig.~\ref{fig:gw_100_graphs}(b).

\section{A Stochastic-Analysis Viewpoint}
\label{sec:an}

Many randomized rounding procedures can be seen from the viewpoint of stochastic analysis: one obtains random unit vectors $u_1,\ldots,u_n\in S^n$ and produces signs $\sigma_1,\ldots,\sigma_n \in \{-1,1\}$. 
In a natural view of \cite{abbasi2020sticky}, the sign is extracted when an associated stochastic process $\{ u_i^T  B(t) \}_{t\ge 0}$ first reaches $\{-1,1\}$,
where $\{B(t)\}_{t\ge 0}$ is a Brownian motion in $\R^n$ adapted to the filtration $\{\mathcal{F}_t\}_{t\ge 0}$.
The corresponding Sticky Brownian Motion
$\forall\, i\in \{1,\ldots,n\}$ is
\begin{equation}\label{eq:defW}
\sigma_i := u_i^T  B(T_i),
\end{equation}
where 
\begin{equation}
T_i := \min\{t\ge 0:\ | u_i^T  B(t) |=1\}.
\end{equation}
This can be extended to ``Slowed-down'' Sticky Brownian Motion \cite{abbasi2020sticky,eldan2019krivine}. Considering first a speed function $\varphi : [-1,1] \to [0, \infty)$ that satisfies
\begin{equation}\label{eq:phi limit assumption}
\lim_{s\to 1^-}\varphi(s)=\lim_{s\to -1^+}\varphi(s)=0
\end{equation}
and 
\begin{equation}
    \forall\, s\in (-1,1),\quad \varphi(s)>0 
\end{equation}
and second a stochastic process $\{W_u^\varphi(t)\}_{(u,t)\in \R^n\times [0,\infty)}$ that satisfies:
\begin{equation}\label{eq:defWS}
\forall(u,t)\in \R^n\times [0,\infty),\quad W_u^\varphi(0) = 0,
\end{equation}
and
\begin{equation}
\mathrm{d} W_u^\varphi(t) = \varphi\big(W_u^\varphi(t)\big) u^T \mathrm{d} B(t),
\end{equation}
one obtains, under mild assumptions \cite{abbasi2020sticky,eldan2019krivine}, 
\begin{equation}
\sigma^\varphi_u := \lim_{t\to\infty} W_u^\varphi(t) \in \{-1,1\} \textrm{ a.s.}
\end{equation}
For the ``Slowed-down'' Sticky Brownian Motion \cite{abbasi2020sticky,eldan2019krivine}, one can show:
\begin{proposition}[Based on \cite{abbasi2020sticky}]
For $u\in \R^n$ and $t\ge 0$ write $W_u^\xi(t)=W_u(t)$ and $\sigma_u^\xi=\sigma_u\in \{-1,1\}$, where 
\begin{equation}\label{eq:def xi}
\forall\, s\in [-1,1],\quad \xi(s) = (1-s^2)^{\alpha},
\end{equation}
for $\alpha>0$. Then,
\begin{equation}\label{eq:subGW}
\forall\, u,v\in S^{d-1},\quad \mathbb{E} \big[\sigma_u \sigma_v \big] \approx 0.878.
\end{equation}
\label{prop:1}
\end{proposition}

We note that the constant in \eqref{eq:subGW} is not exactly the constant of the Goemans-Williamson \cite{Goemans1995} work. (See also Appendix \ref{sec:gw}.)
However, one can consider a different speed-function $\xi$ to obtain the GW constant. 
In particular, by seeing the processes as  Krivine diffusions, one can obtain:
\begin{proposition}[Based on \cite{eldan2019krivine}]
For $u\in \R^n$ and $t\ge 0$ write $W_u^\xi(t)=W_u(t)$ and $\sigma_u^\xi=\sigma_u\in \{-1,1\}$, where 
\begin{equation}\label{eq:defxi2}
\forall\, s\in [-1,1],\quad \xi(s) = \frac{\sqrt{2}}{\sqrt{\pi}} e^{ - \frac12 \Phi^{-1}\big(\frac{1-s}{2}\big)^2  },
\end{equation}
where $\Phi:\R\to \R$ is the standard Gaussian cumulative distribution function, i.e.,
\begin{equation}\label{eq:id}
\forall\,x\in \R,\quad \Phi(x) =  \frac{1}{\sqrt{2\pi}}\int_\infty^x e^{-\frac{s^2}{2}}\,\mathrm{d} s.
\end{equation}
Then,
\begin{equation}
\forall\, u,v\in S^{d-1},\quad \mathbb{E} \big[\sigma_u \sigma_v \big] = \frac{2}{\pi} \arcsin( u^T v).
\end{equation}
\label{prop:2}
\end{proposition}
Compare this to the statement of Proposition~\ref{prop:0}, noting that 
$\arccos(t) + \arcsin(t) = \pi/2$ for
$-1 \le t \le 1$.
The proof relies in seeing the process as discrete-time Krivine diffusions \cite{eldan2019krivine} and applying Theorem 3 of \cite{eldan2019krivine}.

\section{Recursive QAOA\label{sec:rqaoa}}

RQAOA \cite{Bravyi2019} is a recursive algorithm to find the ground state of an Ising Hamiltonian $\hat H_n=\sum_{i,j}J_{i,j}\hat Z_i\hat Z_j + \sum_k J_k \hat Z_k$ with $J_{i,j}, J_k$ as arbitrary real coefficients and $n$ decision variables.
At each step of the recursion a standard QAOA is run to find the state $\ket{\psi^*}=\hat U(\boldsymbol{\beta}^*, \boldsymbol{\gamma}^*)\ket{+}^{\otimes n}$ that minimizes the energy $\braket{\psi^*|\hat H_n |\psi^*}$.
For each edge $(i,j)\in E$ the correlator $\mathcal{M}_{i,j}=\braket{\psi^*|\hat Z_i\hat Z_j |\psi^*}$ is computed.
Next, the decision variable $z_i$ for which $|\mathcal{M}_{ij}|$ is largest is replaced with ${\rm sign}(\mathcal{M}_{i,j})z_j$ to generate a new Ising Hamiltonian $\hat H_{n-1}$ with $n-1$ decision variables.
The recursion stops once the number of variables is below a threshold $n_\text{stop}$.
The remaining problem is solved with a  classical solver.
We refer to Appendix C of \cite{Bravyi2019} for the pseudocode and detailed discussion.

\section{Depth-one RQAOA\label{sec:efficient_rqaoa}}

\begin{figure*}[htbp!]
    \centering
    \includegraphics[width=0.8\textwidth, clip, trim=0 5 0 5]{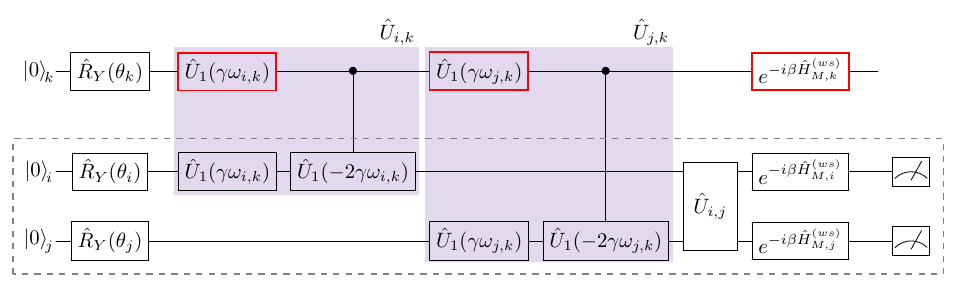}
    \caption{Quantum circuit used to compute the correlator $\langle \hat Z_i\hat Z_j\rangle$.
    The gates highlighted in red do not need to be taken into account.}
    \label{fig:depth_one_rqaoa}
\end{figure*}

Depth-one RQAOA can efficiently be simulated classically \cite{Wang2018}.
Here, we show the algorithm we used to efficiently simulate depth-one WS-RQAOA.
To evaluate the correlator 
$\langle \hat Z_i\hat Z_j\rangle={\rm Tr}\{\rho_{i,j}\hat Z_i\hat Z_j\}$ 
we only need the density matrix $\rho_{i,j}$ of qubits $i$ and $j$, see the circuit in Fig.~\ref{fig:depth_one_rqaoa}.
Qubits $i$ and $j$ are first prepared in the state $(\sqrt{1-c^*_i}\ket{0_i}+\sqrt{c_i^*}\ket{1_i})\otimes(\sqrt{1-c^*_j}\ket{0_j}+\sqrt{c_j^*}\ket{1_j})$.
For each qubit $k\neq i,j$, the cost Hamiltonian applies the gate $\hat U_{i,k}\otimes \hat U_{j,k}$, where $\hat U_{i,k}$ is 
\begin{align}
    \hat U_1(\gamma\omega_{i,k})\otimes \hat U_1(\gamma\omega_{i,k}) \cdot \text{C-Phase}(-2\gamma\omega_{i,k}).
\end{align}
Here, the single-qubit gate $\hat U_1(\phi)$ is ${\rm diag}(1, e^{i\phi})$.
Since the controlled-phase gate $\text{C-Phase}(\phi)={\rm diag} (1,1,1,e^{i\phi})$ commutes with the $\hat U_1$ gate we move all $\hat U_1$ gates to the front of the circuit an apply the phases $\hat U_1(\gamma\sum_{k\neq i,j}\omega_{i,k})$ and $\hat U_1(\gamma\sum_{k\neq i,j}\omega_{j,k})$ to qubits $i$ and $j$, respectively.
Next, we include the effect of the controlled-phase gate of each qubit $k\neq i,j$, initially in the state $\sqrt{1-c_k^*}\ket{0}+\sqrt{c_k^*}\ket{1}$, on the density matrix $\rho_{i,j}$ by computing $(1-c_k^*)\rho_{i,j}+c_k^*\hat U_{ijk}\rho_{i,j} \hat U_{ijk}^\dagger$ where $\hat U_{ijk}={\rm diag}(1, e^{-2i\gamma\omega_{i,k}})\otimes {\rm diag}(1, e^{-2i\gamma\omega_{j,k}})$.
Then we apply the two-qubit operation from the $\omega_{i,j}\hat Z_i\hat Z_j$ term, i.e. $\hat U_{i,j}={\rm diag}(1,e^{i\gamma\omega_{i,j}}, e^{i\gamma\omega_{i,j}}, 1)$, and finally apply the mixer operator before measuring.
This is summarized in Alg.~\ref{alg:rqaoa_p1}.

\begin{algorithm}
 \caption{Depth-one RQAOA}
\label{alg:rqaoa_p1}
\textbf{Initialization:} qubit $i$ and $j$ in state $\ket{0}$.\\
\textbf{Output:} Correlator 
$\langle \hat Z_i\hat Z_j\rangle$ \\ 
\setstretch{1.35}
\SetAlgoLined
 
 Apply $\hat R_Y(\theta_i)$ and $\hat R_Y(\theta_j)$ to qubit $i$ and $j$.
 
 Apply $\hat U_1(\gamma\sum_{k\neq i,j}\omega_{i,k})$ to qubit $i$. \\
 Apply $\hat U_1(\gamma\sum_{k\neq i,j}\omega_{j,k})$ to qubit $j$. \\
 
 \For{$k\neq i,j$}{
    {$\rho_{i,j}\gets(1-c_k^*)\rho_{i,j}+c_k^*\hat U_{ijk}\rho_{i,j} \hat U_{ijk}^\dagger$}
 }
 Apply $\rho_{i,j}\gets \hat U_{i,j}\rho_{i,j}\hat U_{i,j}^\dagger$
 
 Apply mixer $e^{-i\beta(\hat H_{M,i}^{(ws)}\otimes \hat H_{M,j}^{(ws)})}\rho_{i,j}e^{i\beta(\hat H_{M,i}^{(ws)}\otimes \hat H_{M,j}^{(ws)})}$
 
 Measure correlator $\langle \hat Z_i\hat Z_j\rangle ={\rm Tr}\{\rho_{i,j}\hat Z_i\hat Z_j\}$
 
\end{algorithm}

\section{MAXCUT reduction\label{sec:max_cut_reduction}}

Here we show that replacing $\hat Z_i$ by $\pm \hat Z_j$ in a MAXCUT problem results in a new MAXCUT problem with one node less.
Without loss of generality we label the nodes from $1$ to $n$ such that the spin operator $\hat Z_n$ of node $n$ will be replaced by $\hat Z_n=\alpha \hat Z_k$ with $\alpha=\pm1$ and $k<n$.
The MAXCUT Hamiltonian of the weighted graph is
\begin{align}
    \hat H=&\frac{1}{4}\sum_{i,j=1}^n\omega_{i,j}\left(1-\hat Z_i\hat Z_j\right) \\ \notag
     =&\frac{1}{4}\sum_{i,j=1}^{n-1}\omega_{i,j}\left(1-\hat Z_i\hat Z_j\right)+\frac{1}{2}\sum_{i=1}^{n-1}\omega_{i,n}\left(1-\hat Z_i\hat Z_n\right)
\end{align}
We now replace $\hat Z_n=\alpha \hat Z_k$ in the last term 
and, since $\omega_{i,n}\left(1-\alpha \hat Z_i\hat Z_k\right)=\alpha\omega_{i,n}\left(1-\hat Z_i\hat Z_k\right)+\omega_{i,n}(1-\alpha)$, we may write
\begin{align}
   \sum_{i=1}^{n-1}\omega_{i,n}\left(1-\hat Z_i\hat Z_n\right) &= \notag
   \sum_{i=1}^{n-1}\omega_{i,n}(1-\alpha) \\
   &+\sum_{i=1}^{n-1}\alpha\omega_{i,n}\left(1- \hat Z_i\hat Z_k\right).
\end{align}
We neglect the first sum since it is an energy offset that does not affect the optimization.
The Hamiltonian of the reduced problem is therefore
\begin{align} \notag
    \hat H_{n-1}=&\frac{1}{4}\sum_{i,j=1}^{n-1}\omega_{i,j}\left(1-\hat Z_i\hat Z_j\right) \\
    +&\frac{1}{2}\sum_{i=1}^{n-1}\alpha\omega_{i,n}\left(1- \hat Z_i\hat Z_k\right)
\end{align}
This Hamiltonian corresponds to a new graph $E'$ in which the weights $\omega'_{i,j}$ with $i,j=1,...,n-1$ have been updated according to
\begin{align}
    \omega'_{ij} = \left\{
        \begin{array}{cc}
            \omega_{i,j} & \text{if}\quad j\neq k,\\
            \omega_{i,j}+\alpha\omega_{i,n} & \text{if}\quad j= k.
        \end{array}
    \right.
\end{align}

\begin{figure}[htbp!]
    \centering
    \includegraphics[width=\columnwidth]{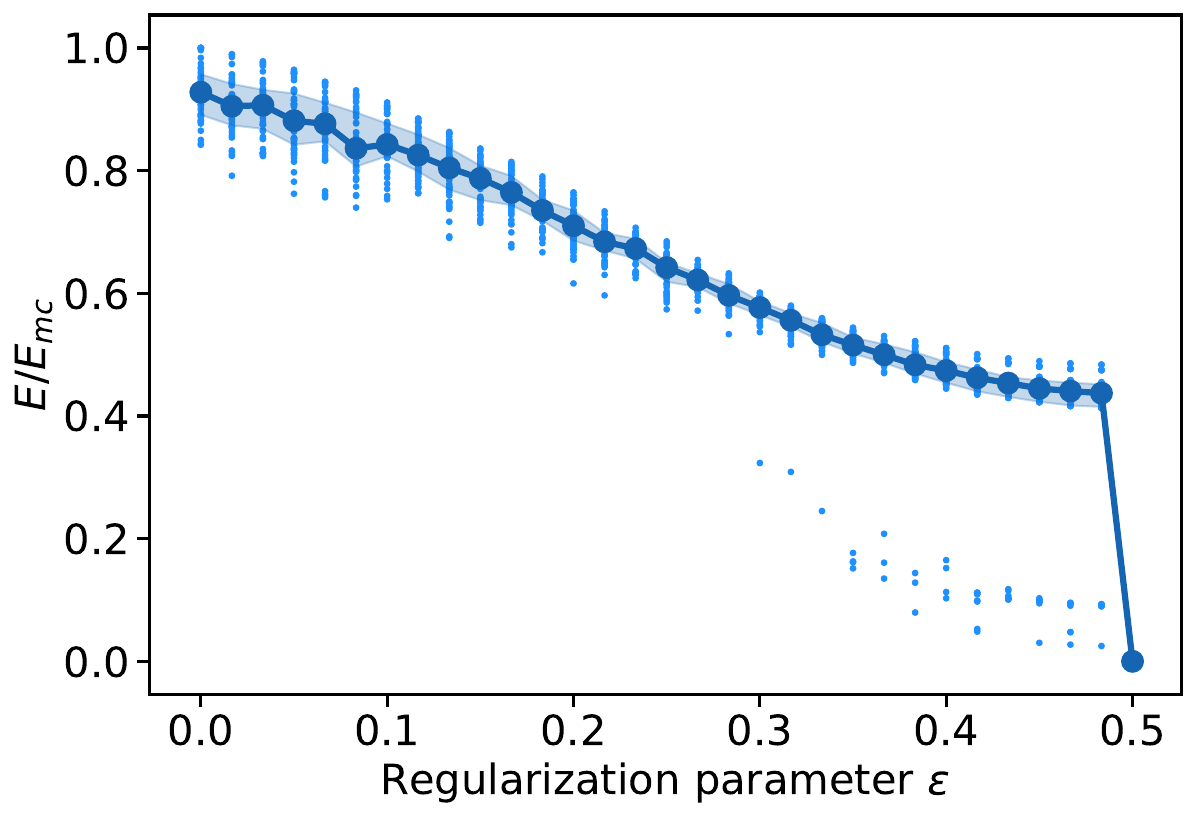}
    \caption{Energy, normalized to the energy of the maximum cut, as a function of $\varepsilon$ for a depth one WS-QAOA. The graphs are the same as those in Fig.~\ref{fig:epsilon_graph} but we used the mixer from Eq.~\eqref{eq:new_mixer} which cannot reproduce the GW at $\varepsilon=0.25$ and depth one.}
    \label{fig:epsilon_ws_mixer}
\end{figure}

\section{Portfolio data\label{sec:gbm}}

The return vectors and covariance matrices used in Sec.~\ref{sec:simulations} are obtained by simulating the price of each asset following a Geometric Brownian motion for $N=250$ days.
The price of asset $i$ on the $k^\text{th}$ day is
\begin{align}
    S_{i,k}=S_{i,0}\exp\left[(\mu_i-\sigma_i^2/2)k/N+\sigma_iW_k\right].
\end{align}
Without loss of generality we set the initial price $S_{i,0}=1$.
We randomly chose each mean $\mu_i$ and standard deviation $\sigma_i$ uniformly form $[-5\%, 5\%]$ and $[-20\%, 20\%]$, respectively.
The Brownian motion is given by $W_k=\sum_{l=0}^j z_l/\sqrt{N}$ where $z_l$ is drawn from the normal distribution.
The return of asset $i$ on the $k^\text{th}$ day is $r_{i,k}=S_{i,k}/S_{i,k-1}-1$.
The average of $r_{i,k}$ gives the mean return of asset $i$ and the covariance of asset $i$ and $j$ is obtained from $r_{i,k}$ and $r_{j,k}$ where $k=1, ..., N$.

\begin{figure*}[htbp!]
    \centering
    \includegraphics[width=\textwidth]{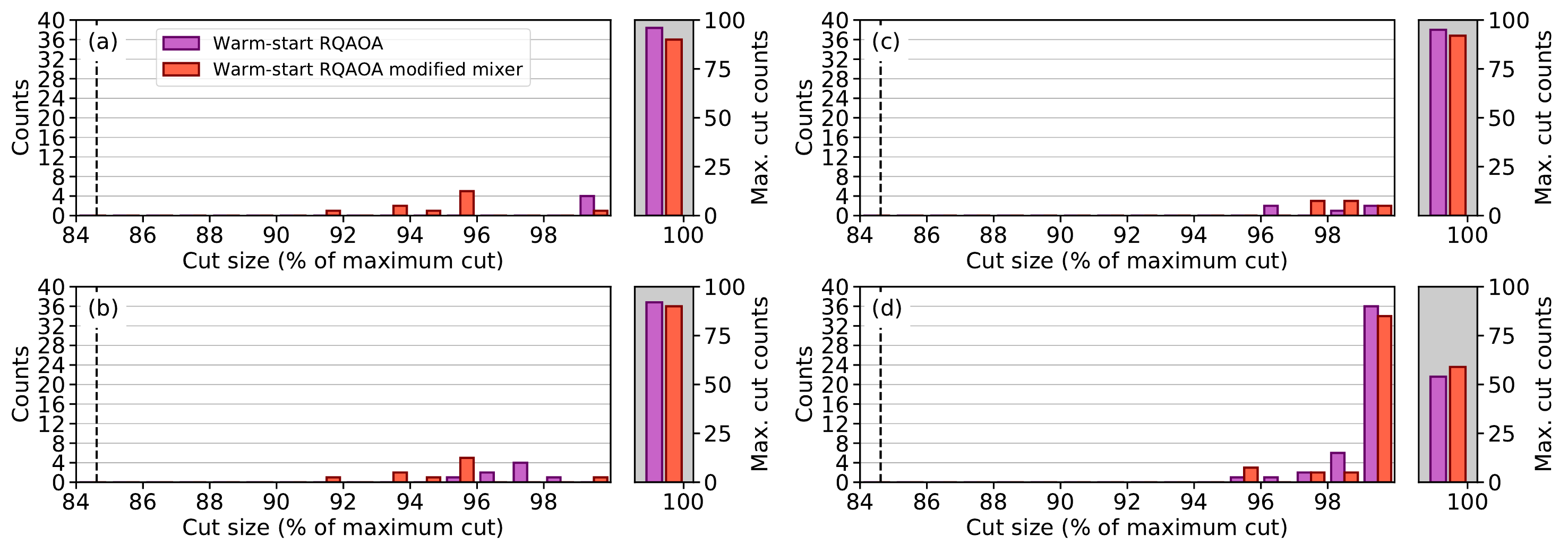}
    \caption{
    Histograms of cut sizes, relative to the maximum cut found by CPLEX, for WS-RQAOA with the modified mixer (red, same data as in Fig.~\ref{fig:max_cut}), and the WS-QAOA with the mixer from Eq.~\eqref{eq:new_mixer}.
    The dashed line shows the hardness factor $11/13$.}
    \label{fig:mixer_comparison}
\end{figure*}

\section{WS-QAOA for MAXCUT with the warm-start mixer\label{sec:ws_mixer_mc}}

In the main text we modified the WS mixer to retain the GW cut.
Here, we explore WS-QAOA for MAXCUT with the mixer of Eq.~(\ref{eq:new_mixer}).
By repeating the analysis done in Fig.~\ref{fig:epsilon_graph} we observe that the energy, normalized to the maximum, decreases as a function of $\varepsilon$ and does not recover at $\varepsilon=0.25$, see Fig.~\ref{fig:epsilon_ws_mixer}.

In addition, we repeat the analysis done in Fig.~\ref{fig:max_cut} but with the mixer of Eq.~\eqref{eq:new_mixer} and compare it to WS-RQAOA with the modified mixer.
Both algorithms have a similar performance, see Fig.~\ref{fig:mixer_comparison}, which suggests that the amount of correlation between the variables at each iteration, e.g. as in Fig.~\ref{fig:epsilon_ws_mixer}, is still sufficient for WS-RQAOA to produce good results even though the GW cut cannot be sampled with certainty at each iteration. 
As the size and complexity of the graphs is increased the performance of WS-RQAOA with the mixer from Eq.~\eqref{eq:new_mixer} decreases compared to WS-RQAOA with the modified mixer and indicates the importance of being able to retain the GW cut.

\bibliographystyle{unsrtnat}
\bibliography{references}
\end{document}